\documentclass{aa}
\usepackage{lscape}
\usepackage{txfonts}
\usepackage{graphicx}
\usepackage{booktabs}
\usepackage{amsmath}
\usepackage{colortbl}
\usepackage{gensymb}
\usepackage{subfig}
\usepackage{rotating}
\usepackage{amssymb}
\usepackage{latexsym}
\usepackage{ifthen}
\usepackage{enumitem}

\usepackage{mathrsfs}
\usepackage{amsfonts}
\usepackage{enumerate}
\newcommand{\asec} {\mbox{$^{\prime \prime}$} }
\newcommand{\amin} {\mbox{$^{\prime}$}}

\def\gsimeq{\hbox{\raise0.5ex\hbox{$>\lower1.06ex\hbox{$\kern-1.07em{\sim}$}$}}} 
\def\lsimeq{\hbox{\raise0.5ex\hbox{$<\lower1.06ex\hbox{$\kern-1.07em{\sim}$}$}}} 

\begin{document}

\title{The  hard X--ray emission of the luminous infrared galaxy  
\\
\textsc{NGC 6240} as observed by {\it NuSTAR}}
\author{S. Puccetti\inst{1,2}, A. Comastri\inst{3}, F. E. Bauer\inst{4,5,6,7},  W. N. Brandt\inst{8,9,10}, F. Fiore\inst{2}, F. A. Harrison\inst{11}, B. Luo\inst{8,9}, D. Stern\inst{12}, C. M. Urry\inst{13}, D. M. Alexander\inst{14}, A. Annuar\inst{14}, P. Ar{\'e}valo\inst{4,15}, M. Balokovi\'c\inst{11}, S. E. Boggs\inst{16}, M. Brightman\inst{11}, F. E. Christensen\inst{17},  W. W. Craig\inst{16,18}, P. Gandhi\inst{14,19}, C. J. Hailey\inst{20},  M. J. Koss\inst{21}, S. La Massa\inst{13}, A. Marinucci\inst{22}, C. Ricci\inst{4,5}, D. J. Walton\inst{12,11}, L. Zappacosta\inst{2}, W. Zhang\inst{23}
}
\authorrunning{S. Puccetti et al. }
\titlerunning{\textsc{NGC 6240}}

   \offprints{S. Puccetti, \email{puccetti@asdc.asi.it} }

\institute{$^1$ASDC--ASI, Via del Politecnico, 00133 Roma, Italy\\
$^2$INAF--Osservatorio Astronomico di Roma, via Frascati
  33, 00078 Monte Porzio Catone (RM), Italy\\
$^3$INAF--Osservatorio Astronomico di Bologna, via Ranzani
1, 40127 Bologna, Italy\\
$^4$EMBIGGEN Anillo, Concepci\'on, Chile\\
$^5$Instituto de Astrof\'{\i}sica, Facultad de F\'{\i}sica, Pontificia Universidad Cat\'{o}lica de Chile, 306, Santiago 22, Chile\\
$^6$Millenium Institute of Astrophysics, Santiago, Chile\\
$^7$Space Science Institute, 4750 Walnut Street, Suite 205, Boulder, Colorado 80301\\
$^8$Department of Astronomy \& Astrophysics, 525 Davey Laboratory, Pennsylvania State University, University Park, PA 16802, USA\\
$^9$Institute for Gravitation and the Cosmos, Pennsylvania State University, University Park, PA 16802, USA \\
$^{10}$Department of Physics, 104 Davey Laboratory, Pennsylvania State University, University Park, PA 16802, USA\\
$^{11}$Cahill Center for Astrophysics, California Institute of Technology, 1216 East California Boulevard, Pasadena, CA 91125, USA\\
$^{12}$Jet Propulsion Laboratory, California Institute of Technology, 4800 Oak Grove Drive, Mail Stop 169-221, Pasadena, CA 91109, USA\\
$^{13}$Yale Center for Astronomy and Astrophysics, Physics Department, Yale University, PO Box 208120, New Haven, CT 06520-8120, USA\\
$^{14}$Department of Physics, Durham University, Durham DH1 3LE, UK\\
$^{15}$Instituto de F\'{\i}sica y Astronom\'{\i}a, Facultad de Ciencias, Universidad de Valpara\'{\i}so, Gran Bretana N 1111, Playa Ancha, Valpara\'{\i}so, Chile \\
$^{16}$Space Sciences Laboratory, University of California, Berkeley CA 94720, USA \\
$^{17}$DTU Space, National Space Institute, Technical University of Denmark, Elektrovej 327, DK-2800 Lyngby, Denmark \\
$^{18}$Lawrence Livermore National Laboratory, Livermore, CA 94550, USA\\
$^{19}$School of Physics \& Astronomy, University of Southampton, Highfield, Southampton SO17 1BJ UK\\
$^{20}$Columbia Astrophysics Laboratory, Columbia University, New York, NY 10027, USA\\
$^{21}$Institute for Astronomy, Department of Physics, ETH Zurich, Wolfgang-Pauli-Strasse 27, CH-8093 Zurich, Switzerland\\
$^{22}$Dipartimento di Matematica e Fisica, Universit\'{a} degli Studi Roma Tre, via della Vasca Navale 84, I-00146 Roma, Italy \\
$^{23}$NASA Goddard Space Flight Center, Greenbelt, MD 20771, USA\\
}

\abstract{We present a broad--band ($\sim$0.3--70~keV) spectral and
  temporal analysis of {\it NuSTAR} observations of the luminous
  infrared galaxy \textsc{NGC 6240}, combined with archival {\it
    Chandra}, {\it XMM--Newton} and {\it BeppoSAX} data. \textsc{NGC
    6240} is a galaxy in a relatively early merger state with two
  distinct nuclei separated by $\sim$1\farcs5. Previous {\it Chandra}
  observations have resolved the two nuclei, showing that they are
  both active and obscured by Compton--thick material. Although they
  cannot be resolved by {\it NuSTAR}, thanks to the unprecedented
  quality of the {\it NuSTAR} data at energies $>$10~keV, we clearly
  detect, for the first time, both the primary and the reflection
  continuum components. The {\it NuSTAR} hard X--ray spectrum is
  dominated by the primary continuum piercing through an absorbing
  column density which is mildly optically thick to Compton scattering
  ($\tau \simeq$ 1.2, $N_{\rm H}\sim 1.5 \times$ 10$^{24}$
  cm$^{-2}$). We detect moderate hard X--ray ($> 10$ keV) flux
  variability up to 20\% on short ($15-20$~ksec) timescales. The
    amplitude of the variability is maximum at $\sim$30 keV and is
  likely to originate from the primary continuum of the southern
  nucleus. Nevertheless, the mean hard X--ray flux on longer
  timescales (years) is relatively constant. Moreover, the two nuclei
  remain Compton--thick, although we find evidence of variability of
  the material along the line of sight with column densities $N_{\rm
    H}$$\leq$2$\times$10$^{23}$ cm$^{-2}$ over long
  ($\sim$3--15~years) timescales. The observed X--ray emission in the
  {\it NuSTAR} energy range is fully consistent with the sum of the
  best--fit models of the spatially resolved {\it Chandra} spectra of
  the two nuclei.  }


\maketitle

\section{Introduction}

Galaxy mergers represent a key phase in most scenarios of galaxy
formation. According to models and simulations (e.g. Springel et
al. 2005, Di Matteo et al. 2005, Hopkins \& Elvis 2010), during the
mergers of massive gas--rich galaxies, cold gas is destabilized and
can rapidly form stars and will ultimately feed the active galactic
nucleus (AGN). The relative importance of starbursts and AGN and the
associated time scales are not known. Disentangling the contribution
of these components (e.g., finding signatures of positive/negative
feedback) in merging systems is crucial for comparing the observations
with model predictions. In this respect \textsc{NGC 6240} plays a key
role as one of the nearest ($z=$0.02448, $D_L$$=$111.2~Mpc,
$D_A$$=$105.9~Mpc) luminous infrared galaxies observed, in a
relatively early merger state (Tacconi et al. 1999, Tezca et al. 2000,
Bush et al. 2008, Engel et al. 2010, Medling et al. 2011). The
infrared luminosity
($L_{IR}\sim10^{11.94}$~$L_\odot$$=$$3.4\times10^{45}$ erg s$^{-1}$,
Wright et al. 1984; Sanders et al. 2003\footnote{$L_{IR}$ is the
  8--1000 $\mu$m luminosity according to the cosmology used in this
  paper.}) is at the boundary between LIRGs (luminous infrared
galaxies: $L_{IR} = (10^{11}-10^{12})$~$L_\odot$) and ULIRGs
(ultra--luminous infrared galaxies: $L_{IR} \ge 10^{12}$~$L_\odot$)
and implies a high star--formation rate (SFR$=61\pm30 M_\odot$
yr$^{-1}$ from Yun \& Carilli 2002, through fitting of the
far--infrared spectral energy distribution). Owing to the SFR which is
high for its mass (e.g. Santini et al. 2009), \textsc{NGC 6240} is
classified as a starburst galaxy.

The optical emission--line spectrum (Fosbury \& Wall 1979, Zasov \&
Karachentsev 1979, Fried \& Schulz 1983, Morris \& Ward 1988, Keel
1990, Heckman et al. 1987, Veilleux et al. 1995, Schmitt et al. 1996,
Rafanelli et al. 1997) is dominated by LINER--like line ratios over
the central $\sim$10~kpc. This spectrum is preferentially associated
to shocks produced by cloud--cloud collisions in the merging system,
rather than to a central photoionizing AGN continuum (see e.g. Fosbury
\& Wall 1979, Fried \& Schulz 1983). Nevertheless, a few
high--excitation features in IR spectra and in optical HST
narrow--band images indicate possible weak AGN signatures in the
southern region of the galaxy (Rafanelli et al. 1997).

The definite presence of an obscured AGN in \textsc{NGC 6240} has
been known since the first {\it ASCA} hard X--ray observations showed
that a reflection spectrum provides the best fit below 10 keV (Iwasawa
\& Comastri 1998).  The primary emission, obscured by a column density
of about $N_{\rm H}$$\sim$2$\times$10$^{24}$ cm$^{-2}$ emerges only at
higher energies and was originally detected by the PDS instrument
onboard {\it BeppoSAX} (Vignati et al. 1999). The nuclear intrinsic
bolometric luminosity is in the quasar regime ($L_{bol}$$\sim 10^{45}$
erg s$^{-1}$, Vignati et al. 1999, Ikebe et al. 2000, Lira et
al. 2002, Boller et al. 2003).
 
The presence of both a transmitted primary absorbed continuum and a
cold reflection component is still under debate. The results
obtained by the broad--band spectral analysis up to energies
of~100--200~keV (Vignati et al. 1999; Ikebe et al. 2000) are
controversial. Vignati et al. (1999) clearly detect the transmitted
primary absorbed continuum, but the presence of the cold reflection
component is not significant. On the other hand, Ikebe et al. (2000),
using {\it RXTE} data, obtained an equally good fit with either a
reflection--dominated spectrum or with a reflected plus transmitted
model. Therefore, the presence of a primary absorbed continuum is not
certain from their spectral analysis. However, Ikebe et al. (2000)
concluded that both components are needed because in the pure
reflection--dominated model the intrinsic power law photon index is
unusually flat ($\Gamma=1.26\pm0.13$).

Historically, \textsc{NGC 6240} has shown moderate X--ray variability
on month/year timescales at energies larger than $\sim$15~keV
(22$\pm7\%$, 54$\pm18$\% and 4$\pm3$\% in the 14--100, 14--24 and
35--100~keV {\it Swift} BAT energy bands, respectively; Soldi et
al. 2013), while at softer energies it has not exhibited significant
variability (Komossa et al. 1998, Netzer et al. 2005).

\textsc{NGC 6240} is the result of merging of two smaller galaxies,
which formed a larger galaxy with two distinct nuclei, with an angular
separation of $\sim$1\farcs5 ($\sim$0.7 kpc). The two nuclei are
observed in near--infrared (e.g. Fried \& Schultz 1983), in the radio
(e.g. Condon et al. 1982) and in optical (e.g. Rafanelli et al. 1997)
images.

At X-ray wavelengths, {\it Chandra}, with its unsurpassed X--ray
angular resolution (FWHM$\sim$0\farcs5), identified that the two
nuclei are both active and both exhibit highly obscured X--ray spectra
typical of Compton--thick AGN (Komossa et al. 2003). The southern
nucleus is also brighter in X--rays than the northern nucleus. The
observed 0.1--10~keV luminosities of the two nuclei are 1.9 $\times
10^{42}$ erg s$^{-1}$ and 0.7 $\times 10^{42}$ erg s$^{-1}$ (Komossa
et al. 2003). The southern nucleus is also brighter based on
3--5~$\mu$m luminosity (see e.g. Mori et al. 2014, Risaliti et
al. 2006).

The black hole mass of the southern nucleus, obtained via high
resolution stellar kinematics, is $(0.84-2.2)\times 10^{9} M_{\odot}$
(Medling et al. 2011). The mass of the northern nucleus, if it follows
the M$_{BH}$$-$$\sigma$ relation (Tremaine et al. 2002), is
$(1.4\pm0.4)\times 10^{8} M_{\odot}$ (Engel et al. 2010). Due to the
large scatter in the relation, this value is subject to systematic
uncertainties much larger than the measurement errors.

The two nuclei are surrounded by extended X--ray emission modelled as
a multi--temperature plasma with temperature and column density
increasing toward the central regions with metal abundances that are
higher than solar ($\frac{Z}{Z_\odot}$$\sim$2--2.5---Netzer et
al. 2005; $\frac{Z}{Z_\odot}$$=$10---Boller et al. 2003; Wang et
al. 2014). The hot gas, associated with highly ionized
\ion{Fe}{XXV}--emitting gas, peaks centrally at the southern nucleus,
with $\sim$30\% of the emission originating outside the nuclear
region. Its temperature, of T$\sim$7$\times$10$^7$~K, indicates the
presence of fast shocks with velocities of the order of $\sim$2200 km
s$^{-1}$ (Feruglio et al. 2013a, Wang et al. 2014). The energetics and
the iron mass in the hot phase suggest that the fast shocks are due to
a starburst--driven wind expanding into the ambient dense
gas. Nevertheless, Wang et al. (2014) were not able to rule out
additional energy injection in the circumnuclear region such as
heating from an AGN outflow. High--resolution mapping of the CO(1--0)
transition (Feruglio et al. 2013b) revealed the presence of a
molecular outflow with velocity $\sim$(200--500) km s$^{-1}$,
originating from the southern nucleus, likely driven by both supernova
winds and radiation from the AGN. Alternatively, the CO(1--0)
molecular gas condenses within the outflowing gas, because the
velocity difference between ionized and molecular outflows would lead
to severe shredding (Veilleux et al. 2013). The coldest and outermost
plasma forms a soft X--ray halo extending up to $\sim$1\farcm5 from
the two nuclei with sub--solar metal abundances
($\frac{Z}{Z_\odot}$$=$0.1--0.5--Nardini et al. 2013).

In this paper, we discuss the broad--band ($\sim$0.3--70~keV) spectral
and temporal properties of {\it NuSTAR} observations of \textsc{NGC
  6240} combined with archival {\it XMM--Newton}, {\it Chandra} and
{\it BeppoSAX} data.  In \S~2, we describe the {\it NuSTAR} and
archival observations. In \S~3 we discuss the variability
properties. Section~4 presents the broad--band spectral analysis, and
in \S~5 we compare the individual {\it Chandra} spectra of the two
nuclei with the {\it NuSTAR} spectrum of the galaxy as a whole. The
results and conclusions are presented in \S~6 and \S~7, respectively.

Throughout this paper, we adopt a $\Lambda$CDM cosmology with
$\Omega_m=0.27$, $\Omega_{\Lambda}=0.73$, and $H_0=67.3$ kms$^{-1}$
Mpc$^{-1}$ (Planck Collaboration 2014).

\section{X--ray observations and data reduction}

\subsection{{\it NuSTAR}}

{\it NuSTAR} consists of two focal plane modules, FPMA and FPMB, is
sensitive at 3--78.4~keV and has a FOV at 10~keV of 10\amin~(Harrison
et al. 2013). The observatory has a 18$\asec$ FWHM with a half--power
diameter of 58$\asec$. We analyzed the {\it NuSTAR} observation of
\textsc{NGC 6240} performed in March 2014. The observation log is
given in Table~\ref{tablog}.

\begin{table*}
\footnotesize
\caption{{\it NuSTAR} \textsc{NGC 6240} observation log}
\centering
\begin{tabular}{lcccccccc}
\hline
Observation ID$^a$ & RA\_PNT$^b$ & DEC\_PNT$^c$ &  Exposure$^d$ & Start Date$^e$ & rate$^f$& background$^g$ \\ 
 & (deg.) & (deg.) & (ksec) &  &  (cts/s) & \\ 
\hline
 60002040002 &  253.2528  & 2.4308  &  30.8  & 2014-03-30T13:41:07 & 0.159$\pm$0.002 & 7.6\%\\ [1ex]
\hline 
\end{tabular}

$^a$Observation identification number; $^b$Right Ascension of the
pointing; $^c$declination of the pointing; $^d$ total net exposure
time; $^e$start date and time of the observation;
$^f$mean value of the net count rate in the circular source extraction
region with 75$\asec$ radius in the energy range 3--78.4 keV; $^g$
background percentage in the circular source extraction region with
75$\asec$ radius and in the energy range 3--78.4 keV.
\label{tablog}
\end{table*}

\begin{figure}
\begin{center}
\includegraphics[width=6.7cm]{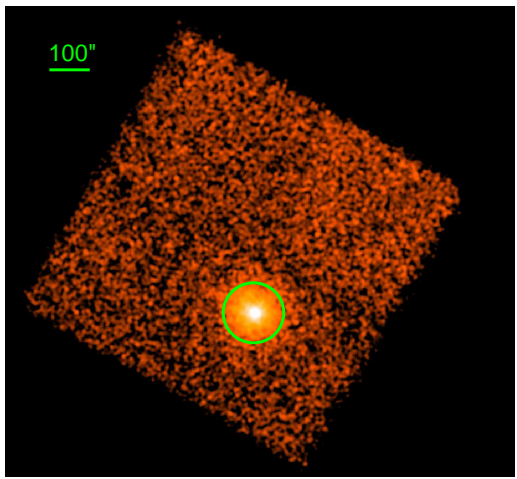}
\caption{3--78.4~keV {\it NuSTAR} image for the FPMA module. The image
  was smoothed with a Gaussian filter with $\sigma=1.5$~pixels ($\sim
  3\farcs6$). The green circle is centered on the peak of the emission
  and has a radius of 75$\asec$.  }
\label{imanu}
\end{center}
\end{figure}

The raw events files were processed using the {\it NuSTAR} Data
Analysis Software package v. 1.4.1
(NuSTARDAS).\footnote{http://heasarc.gsfc.nasa.gov/docs/nustar/analysis/nustar\_swguide.pdf}
Calibrated and cleaned event files were produced using the calibration
files in the {\it NuSTAR} CALDB (20150225) and standard filtering
criteria with the {\it nupipeline} task.  We used {\it nuproducts}
task included in the NuSTARDAS package to extract the {\it NuSTAR}
source and background spectra using the appropriate response and
ancillary files. We extracted spectra and light curves in each focal
plane module (FPMA and FPMB) using circular apertures of radius
75$\asec$, corresponding to $\sim 80\%$ of the encircled energy,
centered on the peak of the emission in the 3--78.4~keV data (see
Fig. \ref{imanu}). Background spectra were extracted using
source--free regions on the same detector as the source. As shown in
Table~\ref{tablog}, the background count--rates are a low fraction ($<
8\%$) of the source count--rates.

The spectra were binned according to two criteria: (i) following the
energy resolution multiplied by a factor $\sim$0.4 at all energies,
when possible; (ii) requiring a signal--to--noise ratio $> 4.5$.
Spectral fitting was performed in the {\it NuSTAR} band alone and
simultaneously with lower energy X--ray data (see below).

\subsection{ {\it Chandra}}

{\it Chandra}, with its unsurpassed X--ray angular resolution
(FWHM$\sim$0\farcs5), is an excellent complement to {\it NuSTAR} for
analyzing the two nuclei and nearby regions and for studying possible
contamination from sources that cannot be resolved by {\it NuSTAR}.

We analyzed four archival {\it Chandra} observations collected in 2001
and 2011 (see Table~\ref{tabarchive}) using the {\it Chandra}
Interactive Analysis Observations (CIAO) software (v4.5; Fruscione et
al. 2006) and the standard data reduction procedures. The {\it
  specextract} task was used to extract the spectra. The background
spectra were extracted from source free regions at distances greater
than about 1\farcm5 from the nuclei of the galaxy in order to avoid
contamination from the X--ray diffuse emission, which is still seen on
scales of 1\farcm~(see Nardini et al. 2013). The spectra were binned
to have at least 25 total counts per bin. The nuclei showed negligible
pile--up (less than a few $\%$). For the final spectral analysis, we
used only the two ACIS--S observations (2001 and 2011) due to
calibration uncertainties above 5~keV of HETG 0th order ACIS--S
spectra.

To avoid a loss of angular sensitivity, we registered the astrometry
of the ACIS--S 2001 data to the ACIS--S 2011 data, using the X--ray
positions of bright common point--like sources. We found that the
ACIS--S 2001 right ascension astrometric correction is --0\farcs1,
while the declination astrometric correction is negligible
(i.e. 0\farcs01).

\begin{figure}
\centering
\includegraphics[width=6.2cm]{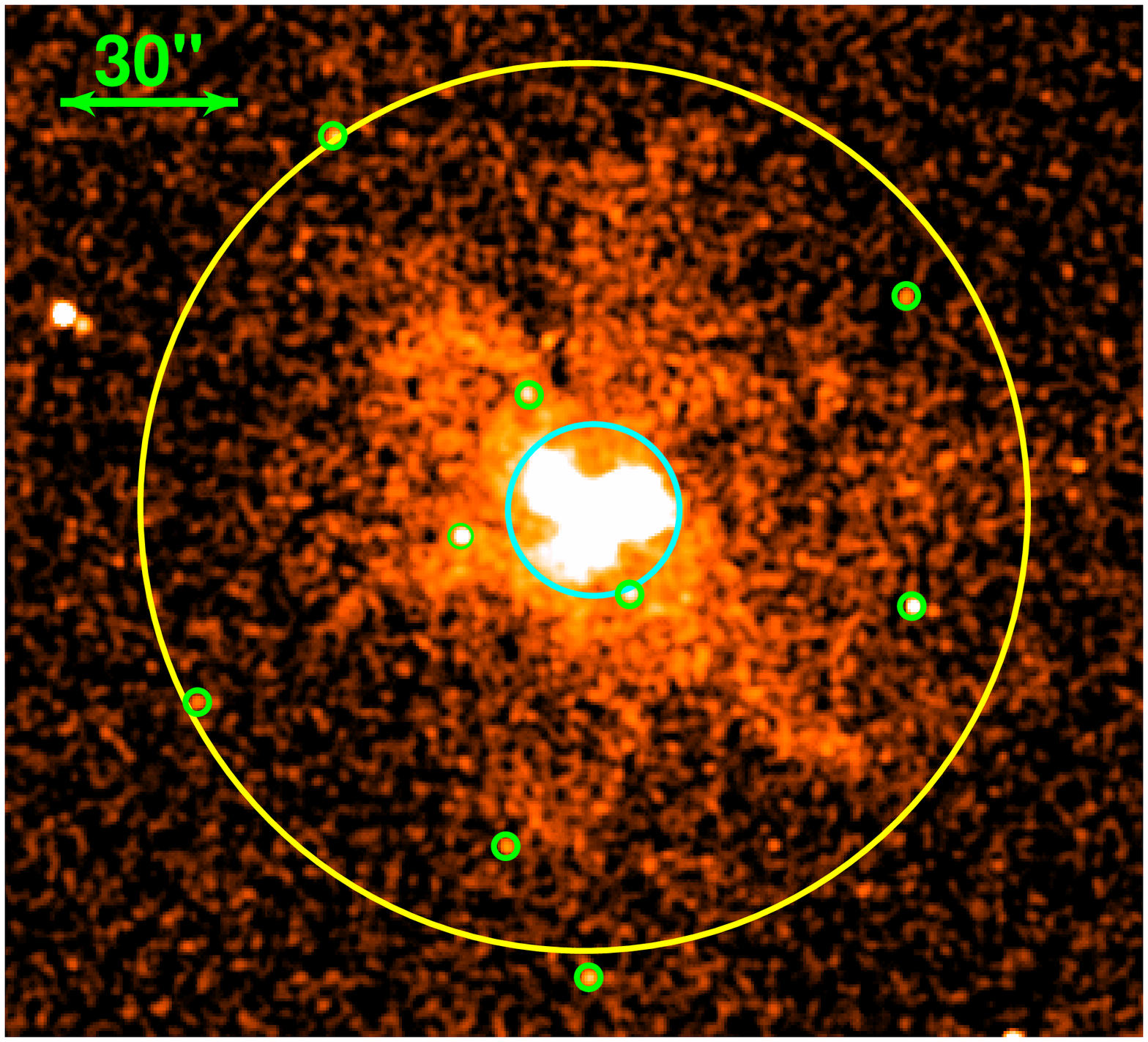}
\includegraphics[width=6.5cm]{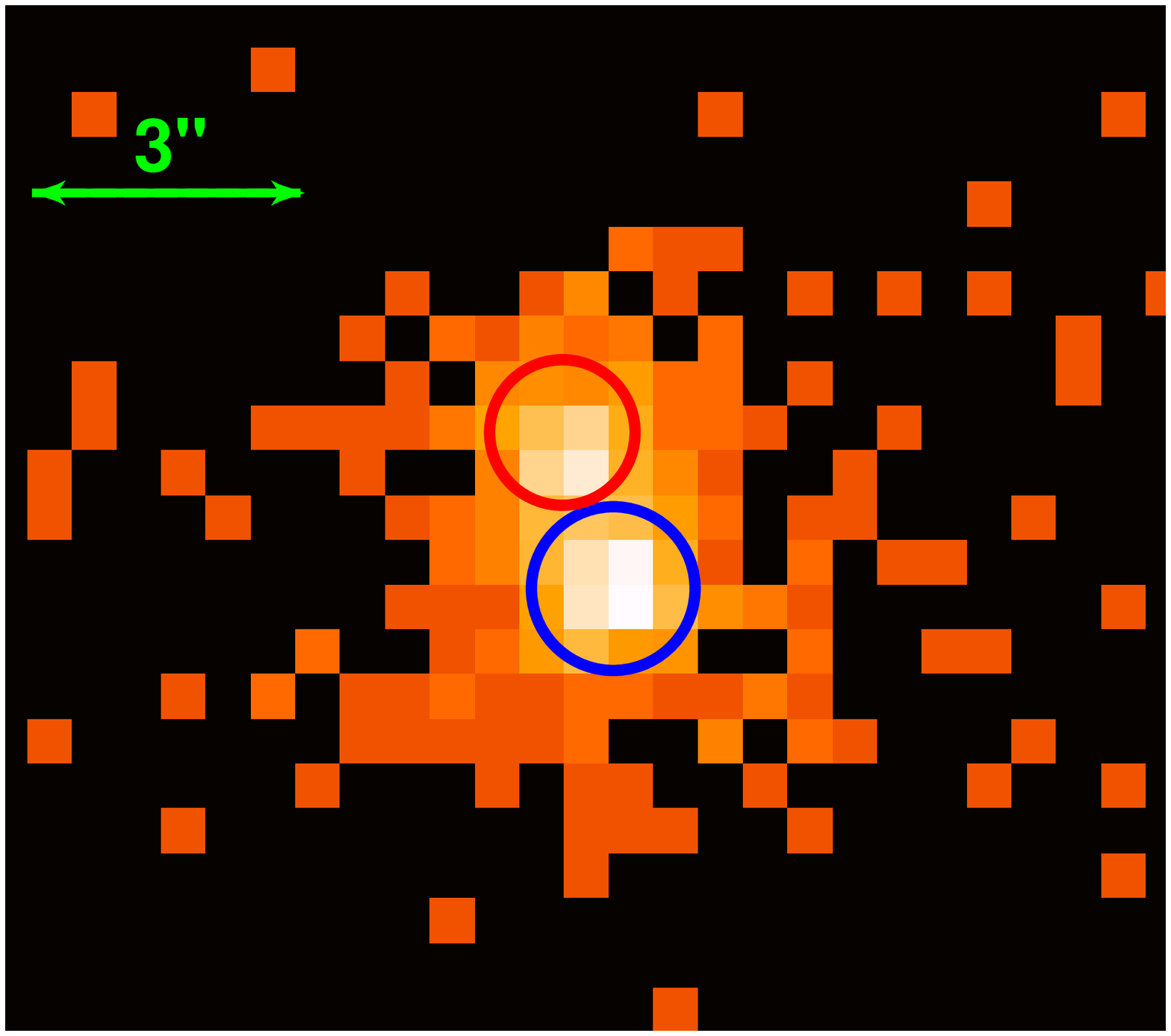}
\includegraphics[width=6.9cm]{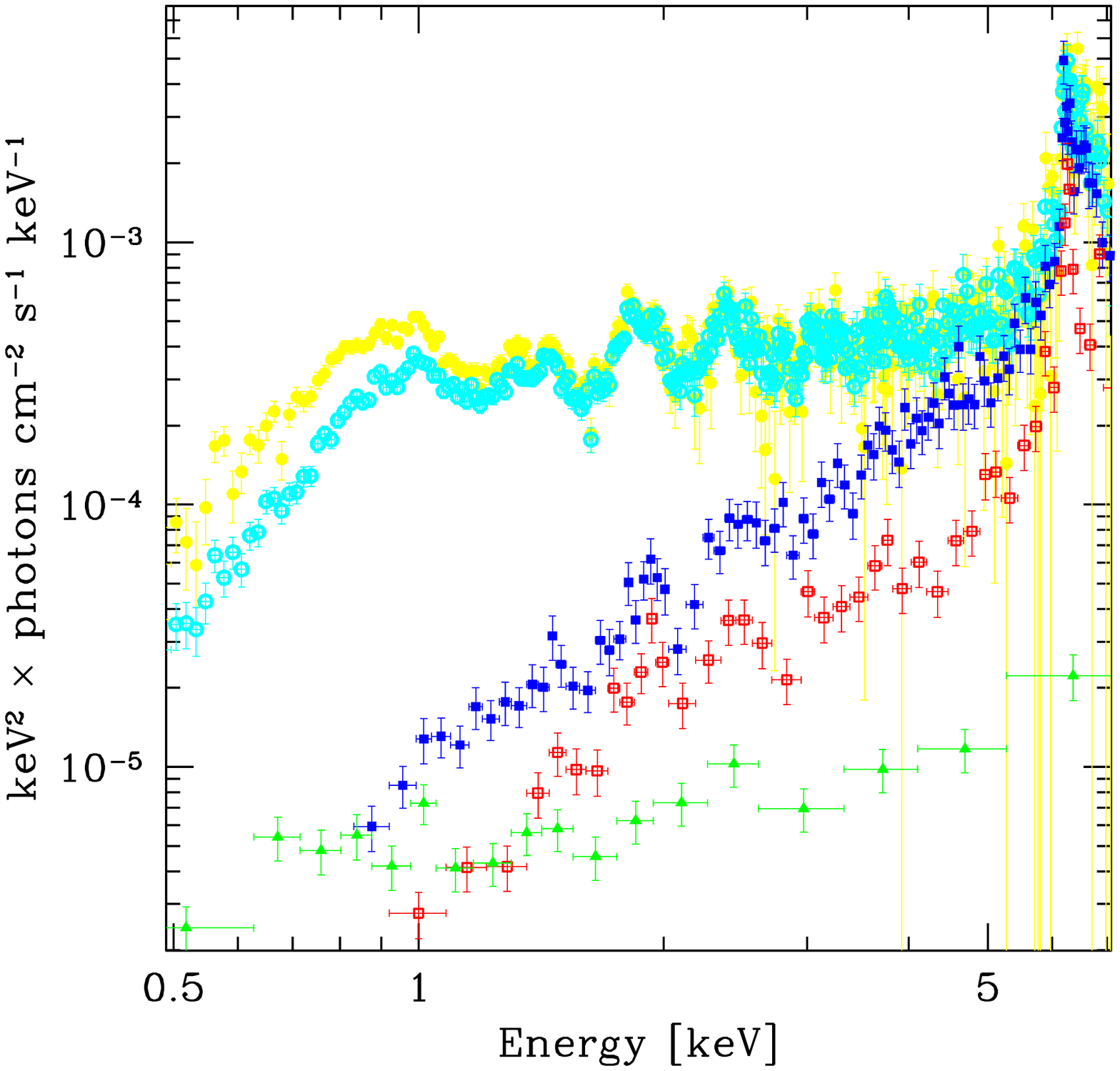}
\caption{\small {{\it Upper panel}: 0.3--8~keV image mosaic of the 2001 and
  2011 {\it Chandra} observations (see Table~\ref{tabarchive}). The
  image was smoothed with a Gaussian filter with $\sigma=1.5$~pixels
  ($\sim 0\farcs75$). The big yellow circle (75$\asec$ radius) marks
  the {\it NuSTAR} source extraction region; the cyan circle
  (14\farcs5 radius) marks the {\it Chandra} source extraction region;
  the small green circles (2\asec~radius) mark the 9 serendipitous
  point sources with S/N $\ge 2$ in the 0.3--8 keV band within/very
  near the area corresponding to the {\it NuSTAR} extraction
  region. {\it Middle panel}: 6--6.5~keV image mosaic of the two {\it
    Chandra} observations: zoomed into the center of the galaxy: the
  southern nucleus and northern nucleus are marked by blue and red
  circles, respectively. {\it Lower panel}: {\it Chandra} spectra,
  unfolded with the instrument response, of the observation 12713 (see
  Table~\ref{tabarchive}). The yellow, cyan, blue and red spectra are
  extracted in the regions marked with the same color in the upper and
  middle panels. The green spectrum represents the ``contamination''
  spectrum by the nearby bright sources detected in the {\it NuSTAR}
  source extraction region. This ``contamination'' is negligible in the {\it
    NuSTAR} extraction region, whose spectrum is shown in yellow.}}
\label{imacha}
\end{figure}

From the analysis of the {\it Chandra} images (see upper panel of
Fig. \ref{imacha}), using an aperture photometry technique (see
e.g. Puccetti et al. 2009), we identified 9 serendipitous point
sources with S/N $\ge 2$ in the 0.3--8 keV band within/very near the
area corresponding to the {\it NuSTAR} extraction region (i.e. a
circle with a radius of 75$\asec$). We co-added the source spectra,
background spectra and ancillary files of the nine sources with {\it
  addascaspec} FTOOLS v.6.13, leaving the original exposure time of a
single source in the final co--added spectrum. The total 0.3--8~keV
flux of these 9 sources (green spectrum in the lower panel of
Fig. \ref{imacha}) is definitely negligible in comparison to the flux
in the {\it NuSTAR} extraction region (yellow spectrum in the lower
panel of Fig. \ref{imacha}); it is fitted by a black body of
$T\sim$0.8~keV and there is no evidence of any iron line
emission. Therefore, the co--added spectrum of the nine brightest
``contaminants'' is consistent with a typical spectrum of
ultra--luminous X-ray sources (ULXs), which turns over very quickly at
about 10~keV as discovered by {\it NuSTAR} (see e.g. Walton et
al. 2015). For these reasons, we conclude that any contamination by
close point--like sources in the {\it NuSTAR} spectrum is negligible.

The second step was to extract the {\it Chandra} spectra of the whole
galaxy (i.e. combination of the two nuclei, extended emission and
serendipitous sources) using a circular region of radius 14\farcs5 to
maximize the S/N of the spectra. In this way, we exclude most of the
very soft X--ray emission in the halo of \textsc{NGC 6240} ($\leq
2$~keV, see Nardini et al. 2013). In any case, this component is not
detectable in the energy band covered by {\it NuSTAR}, which has a low
energy limit at 3 keV.  This is clearly seen in the comparison of the
cyan and yellow spectra in the lower panel of Fig. \ref{imacha}.

\begin{figure*}
\centering
\includegraphics[width=7.2cm,height=6.7cm]{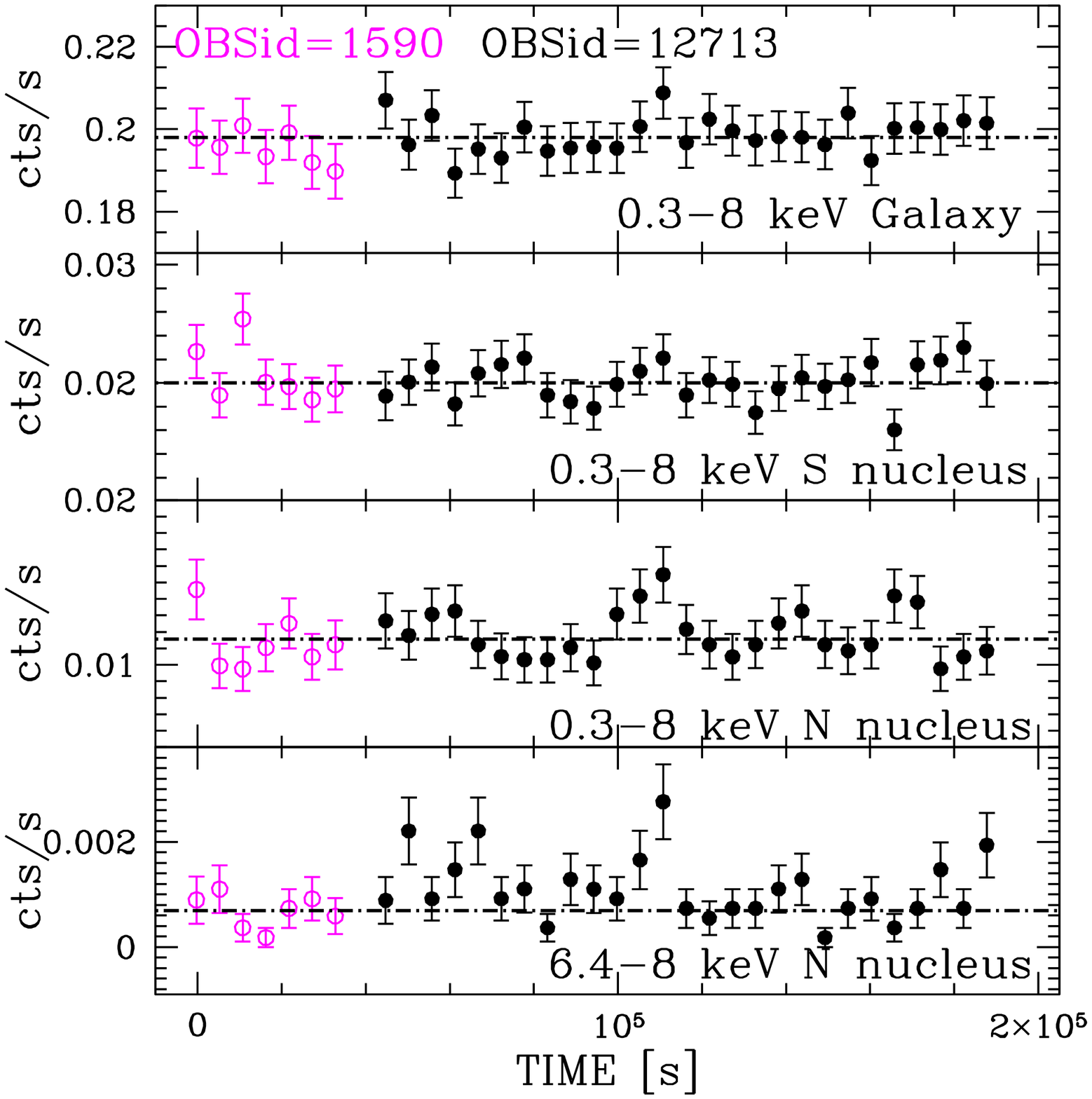}
\includegraphics[width=7.2cm,height=6.7cm]{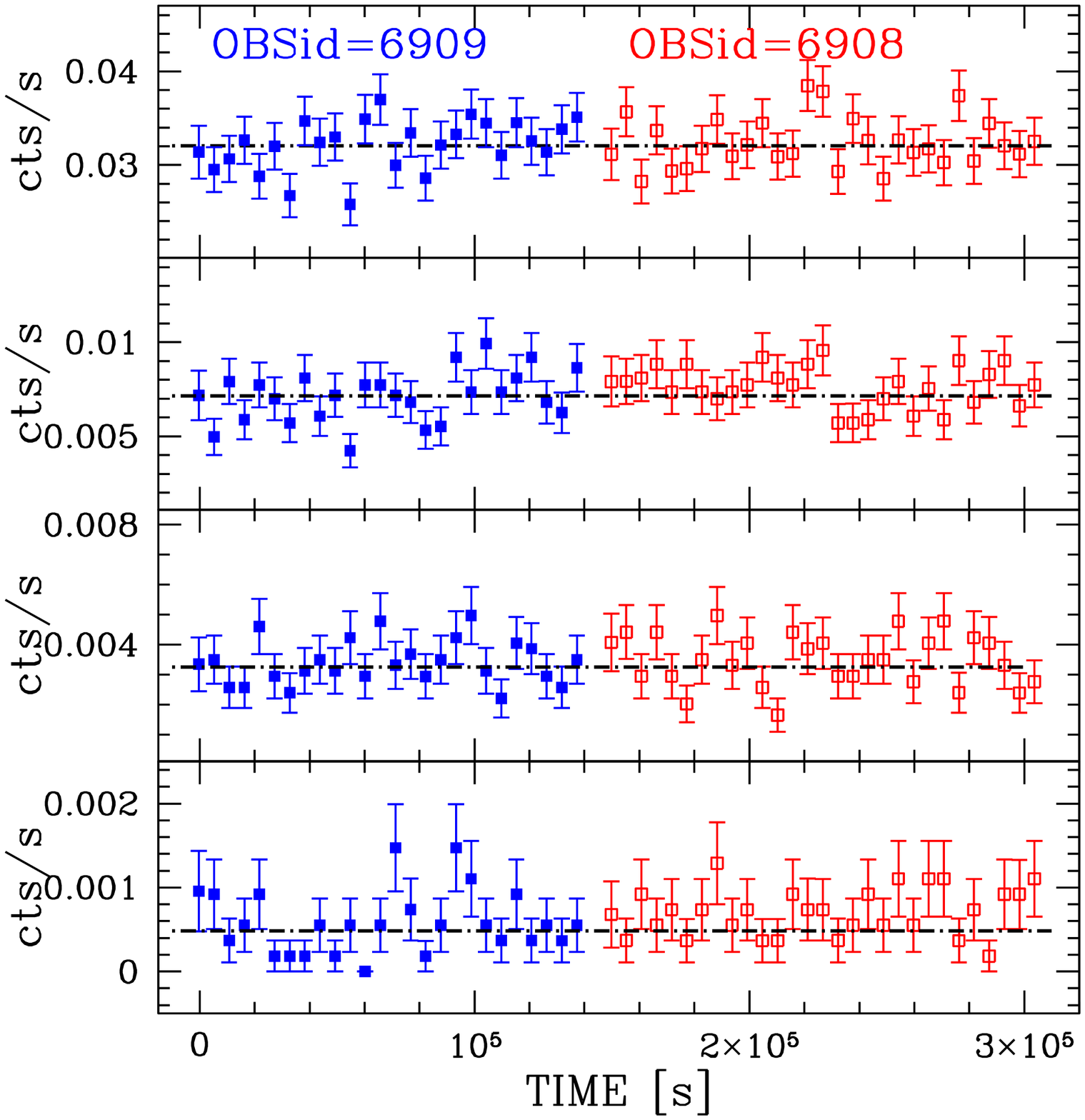}
\caption{{\it Chandra} background--subtracted light curves in
    bins of 5500 s. From top to bottom panel: 0.3--8~keV count rates
    of the whole \textsc{NGC 6240} galaxy, 0.3--8~keV count rates of
    the southern nucleus, 0.3--8~keV count rates of the northern
    nucleus and 6.4--8~keV count rates of the northern nucleus. The
    dot--dashed lines correspond to the values of the count rate
    weighted means. {\it Left panels} show the light curves of the two
    ACIS--S {\it Chandra} observations (see Table~\ref{tabarchive}):
    1590 (magenta open dots) and 12713 (black solid dots). {\it Right
      panels} show the light curves of the zero--order ACIS--S HETG
    {\it Chandra} data: 6909 (blue solid squares) and 6908 (red open
    squares). Note that the x$-$axis does not represent the real
    time.}
\label{lccha}
\end{figure*}

\begin{figure}
\begin{center}
\includegraphics[width=6.7cm]{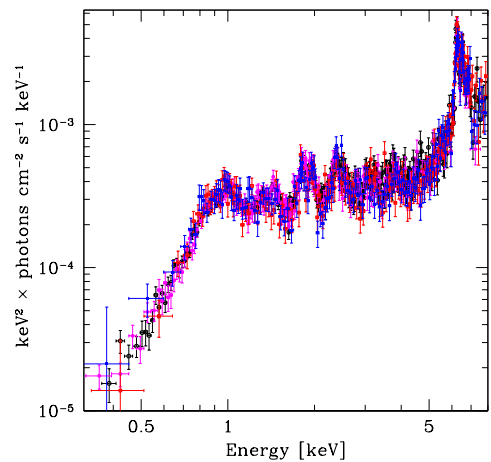}
\caption{{\it Chandra} spectra of the whole \textsc{NGC 6240} galaxy,
  unfolded with the instrument response, extracted in the cyan region
  in the upper panel of Fig. \ref{imacha}. The colors represent the
  four observations (see Table~\ref{tabarchive}). Since no significant
  spectral variability was present, we co--added the ACIS--S spectra.
}
\label{specha}
\end{center}
\end{figure}

Finally, we verified that the whole galaxy did not show significant
variations during the four {\it Chandra} observations (see
Fig. \ref{lccha}). We used the $\chi^2$ statistics with a threshold
probability of $2\%$ to test the hypothesis of constant light
curves. The 0.3--8~keV light curve of the whole galaxy is consistent
with being constant between the two ACIS--S and the two ACIS--S HETG
observations. The 6.4--8~keV count rates are only $\sim$0.3\% and
$\sim$10\% of those in the full energy band, for the ACIS--S and the
zero--order ACIS--S HETG, respectively. For this reason small
amplitude hard X--ray variability is difficult to be detect in the
full energy range; therefore we also verified that the 6.4-8 ~keV
count rates are statistically constant. The ACIS--S and ACIS--S HETG
spectra are fully consistent each other (see Fig. \ref{specha})
confirming that the variability of the whole galaxy is not
significant. Therefore in order to improve the statistics, for the
final spectral analysis we used the co--added spectra of the two
ACIS--S (2001 and 2011) observations, and we also combined the
corresponding background spectra, response and ancillary files. We
used {\it addascaspec} FTOOLS v.6.13, which combines spectra and
normalized backgrounds according to the method explained in the {\it
  ASCA} ABC
guide.\footnote{http://heasarc.gsfc.nasa.gov/docs/asca/abc/}

In addition to the spectrum of the whole galaxy, we also used
the ACIS--S spectra of each of the two individual nuclei (see Komossa et
al. 2003). The spectra were extracted using circular regions centered
on R.A.$=$16:52:58.896, Dec.$=$+02:24:03.36 and R.A.$=$16:52:58.922,
Dec.$=$+02:24:05.03 and with a radius of 0$\farcs$9 for the southern
nucleus (the brightest one) and 0$\farcs$8 for the northern nucleus,
respectively (blue and red circles and spectra in the middle and
bottom panel of Fig. \ref{imacha}, respectively).

The 0.3-8~keV light curves of the southern and northern nucleus are
consistent with being constant between the two ACIS--S and the two
ACIS--S HETG observations (see Fig. \ref{lccha}). We also searched for
variability in the two sub--bands: 0.3-6.4~keV and 6.4-8~keV. We found
variability only in the 6.4-8~keV light curve of the northern nucleus
in the ACIS--S 2011 observation ($\chi^{2}_{\nu}$$=$~1.84 with
probability of false positive $\le 0.6\%$, see bottom panels of
Fig. \ref{lccha}). Nevertheless the count rate variations are small
and the median values of the count rates of the two ACIS--S
observations are fully consistent (i.e. (0.011$\pm$0.001)
cts/s). Therefore we used also for the nuclei the co--added spectra of
the two ACIS--S (2001 and 2011) observations. We stress that the
0.3-8~light curve of the northern nucleus of ACIS--S 2011 shows a
systematic residual respect to a fit with a constant in the time
interval 100--120~ksec in Fig. \ref{lccha}, which could be due to the
6.4--8~keV variability.

\begin{table*}
\footnotesize
\caption{ \textsc{NGC 6240} archival data}
\centering
\begin{tabular}{lccccccccc}
\hline
Observatory & Observation ID$^a$ &  Exposure$^b$ & Start Date$^c$ & & & &ref$^d$\\
 & & (ksec) & \\
\hline
 {\it BeppoSAX} & 5047600100 & 120 &    1998-08-14  & & & & 1 \\
 {\it Chandra} ACIS--S & 1590 & 36.7 &   2001-07-29T05:44:22 & & & & 2,3,4 \\
 {\it Chandra} ACIS--S HETG & 6909 & 141.2 &  2006-05-11T01:41:48 &  && & 3,4,5 \\
 {\it Chandra} ACIS--S HETG & 6908 & 157.0 &  2006-05-16T11:22:54 &  & & & 3,4,5 \\
 {\it Chandra} ACIS--S & 12713 & 145.4 &  2011-05-31T04:15:57  & & & & 3,4 \\
 &&   && PN$^e$& MOS1$^e$& MOS2$^e$ &\\
 {\it XMM--Newton} &  0101640101 &30.1&  2000-09-22T01:38:46  & 11.3 & 15.4 &15.4& 6,7\\
 {\it XMM--Newton} &  0101640601 &19.0&  2002-04-12T21:37:24  & 5.3&10.2 &10.0 & 7\\
 {\it XMM--Newton} &  0147420201 &31.6&  2003-04-14T18:06:14  & 3.4 & 7.1 & 7.9& 7 \\
 {\it XMM--Newton} &  0147420301 &28.1&  2003-04-18T21:01:56  & 0 & 0 & 0 & 7 \\
 {\it XMM--Newton} &  0147420401 &14.1&  2003-08-13T10:29:32  & 7.0 & 10.8 & 10.7& 7 \\
 {\it XMM--Newton} &  0147420501 &31.2&  2003-08-21T10:25:09  & 3.4 & 1.7& 1.8& 7 \\
 {\it XMM--Newton} &  0147420601 & 9.2&  2003-08-29T11:25:15  &1.4 &1.8&1.9&   7 \\

\hline 
\end{tabular}
  
$^a$ Observation identification number; $^b$ total net exposure time;
$^c$ start date and time of the observation interval; $^d$ reference:
1: Vignati et al. (1999), 2: Komossa et al. (2003), 3: Nardini et al. (2013), 4: Wang et
al. (2014), 5: Shu et al. (2011), 6: Boller et al. (2003), 7: Netzer et al. (2005) ;
$^e$ Good Time Interval after background cleaning.
\label{tabarchive}
\end{table*}

\subsection{ {\it XMM--Newton}}

In the spectral analysis we also considered six {\it XMM--Newton}
observations with a non--zero exposure time after background cleaning,
obtained between September 2000 and August 2003 (see
Table~\ref{tabarchive}).  The high spectral resolution of {\it
  XMM--Newton} at $\sim$6 keV and the good counting statistics of the
archival observations allow us to constrain the spectral shape and the
intensity of the iron line complex. We reduced the data using the
Science Analysis Software (SAS ver. 14.0.0), following the standard
method described in the data analysis threads and the ABC Guide to
{\it XMM--Newton} Data
Analysis.\footnote{http://xmm.esac.esa.int/sas/current/documentation/threads/}
Epochs of high background events were excluded; only events
corresponding to pattern 0--12 for MOS1/MOS2 and pattern 0--4 for PN
were used. From the analysis of the {\it Chandra} spectra, we found
that the AGN and hot gas in the galaxy are extended less than
  15\asec, thus the spectra were extracted using a circular region of
radius 40$\asec$, which corresponds to $\sim 90\%$ of the encircled
energy, similar to the {\it NuSTAR} data, which includes all the galaxy
emission at energy $\geq 2$~keV. Extraction of the background spectra
and binning were done similar to the {\it NuSTAR} analysis.

\begin{figure}
\begin{center}
\includegraphics[width=6.7cm]{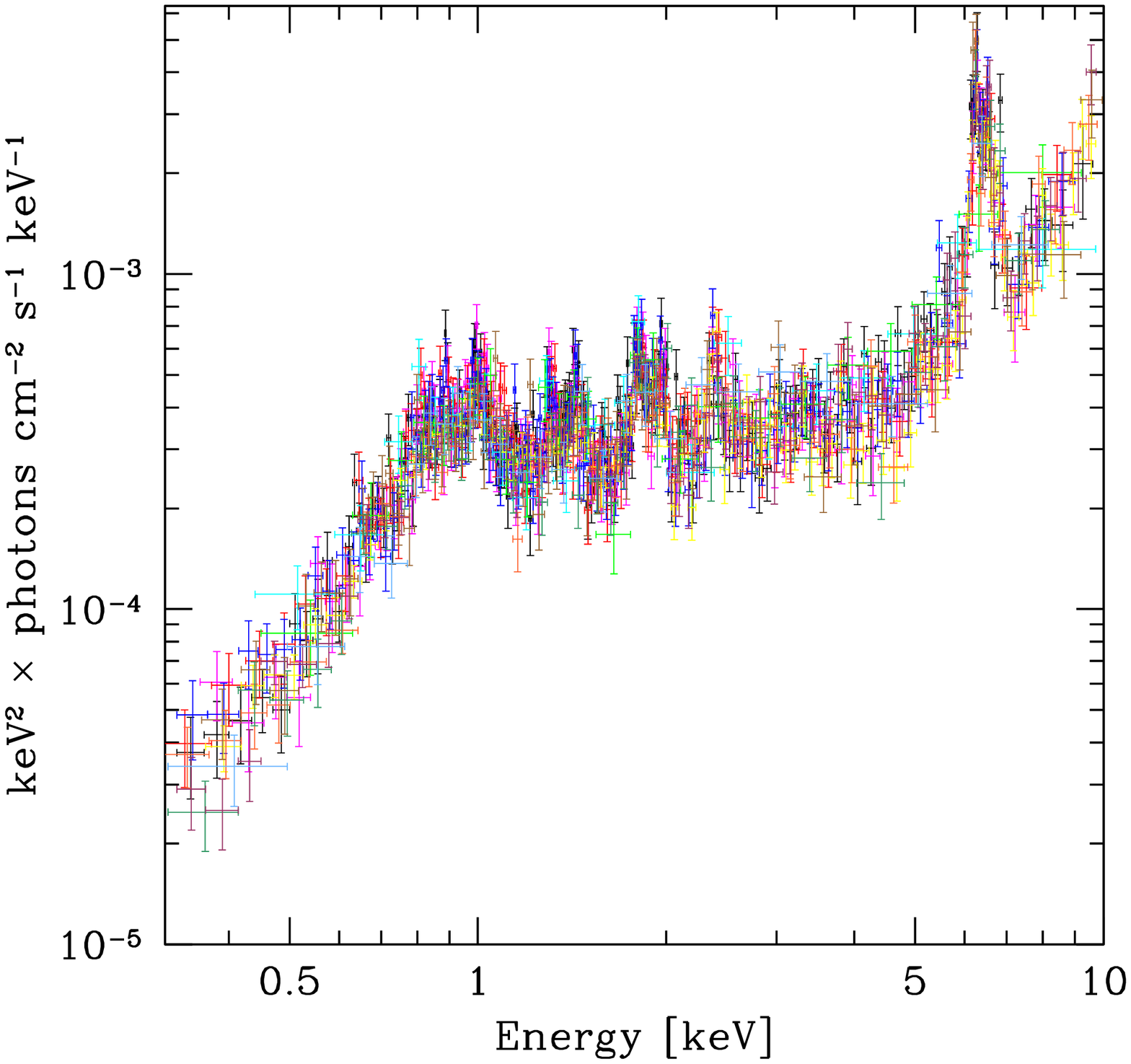}
\caption{ {\it XMM--Newton} spectra of the whole \textsc{NGC 6240}
  galaxy, unfolded with the instrument response. The colors represent
  the EPIC--PN and co-added EPIC--MOS1 and EPIC--MOS2 spectra for all
  the observations (see Table~\ref{tabarchive}).  }
\label{spexmm}
\end{center}
\end{figure}

The whole \textsc{NGC 6240} galaxy, as already pointed out by Netzer
et al. (2005), does not exhibit significant spectral variability
during the {\it XMM--Newton} observations (see also
Fig. \ref{spexmm}). Therefore, to improve the statistics, we
considered the combination of all MOS1 and MOS2 spectra, and of the
individual PN spectra. In the spectral analysis we used the {\it
  XMM--Newton} data between 2--10~keV only, to avoid a larger soft
X--ray contribution in comparison with the {\it Chandra} spectra.

\subsection{{\it BeppoSAX} PDS data}

\textsc{NGC 6240} was observed by {\it BeppoSAX} (Boella et al. 1997)
on 1998 August 14 (Vignati et al. 1999) with the LECS (Low Energy
Concentrator Spectrometer; 0.45--4~keV), the two MECS (Medium Energy
Concentrator Spectrometers; 1.65--10.5~keV) and the PDS (Phoswitch
Detector System; 15--200~keV, Frontera et al. 1997) (see
Table~\ref{tabarchive}). Data products were retrieved from the {\it
  BeppoSAX} archive at the ASI Science Data
Center.\footnote{http://www.asdc.asi.it/mmia/index.php?mission=saxnfi}
The data were calibrated and cleaned using \textsc{SAXDAS}
software. The MECS/LECS event files were screened adopting standard
pipeline selection parameters, and the spectra were extracted from
4\amin~radii apertures, corresponding to $\sim 90\%$ of the encircled
energy. The two MECS units were combined after renormalizing to the
MECS1. The PDS is a collimator instrument with a field of view of
$\sim1^{\circ}$.3 (FWHM). The PDS data were calibrated and cleaned
following the standard ``fixed Rise Time threshold'' method for
background rejection. Following the method used for the {\it NuSTAR}
data, we estimate that for the MECS data, the contamination by the
nearby bright sources is less than $10\%$. We do not use the LECS data
to avoid a larger soft X--ray contribution in comparison with data
from other instruments used in this work.

\section{Light curve variability}

With a few exceptions (e.g. the Compton--thick to Compton--thin
changes in \textsc{NGC 1365}, Risaliti et al. 2005, 2007), variability
in Compton--thick AGN is extremely rare at energies
$<10$~keV. Constraining intrinsic variability in such obscured sources
requires statistically good sampling of the 30--50 keV energy range,
where the primary continuum can arise. The analysis of the light curve
in different energy ranges returns raw information on the origin of
the variability (i.e., variations in Eddington ratio, column density,
photon index of the primary continuum, reflection component). When
data with good statistics are available, the following time/count rate
resolved spectral analysis allow us to constrain better the variable
spectral parameters, and place constraints on the geometry of the
reflecting/absorbing medium (see e.g. \textsc{NGC 4945} Puccetti et
al. 2014; \textsc{NGC 1068} Marinucci et al. submitted). Unfortunately
for \textsc{NGC 6240} the statistics do not allow this detailed
spectral analysis.

The {\it NuSTAR} \textsc{NGC 6240} observation lasted $\sim 60$ ksec,
but the Earth occults the target for approximately half the orbit,
yielding a net exposure time of 30.8~ksec. We analyzed the {\it
  NuSTAR} light curves in bins of 5800 s ($\sim 1$ satellite
orbit). We used the $\chi^2$ and the Kolmogorov Smirnov tests to
verify the constancy of the light curves. A light curve is considered
definitely variable if the probability of false positive is $\le 2\%$
for at least one of the two tests.

\begin{figure}
\includegraphics[width=8cm]{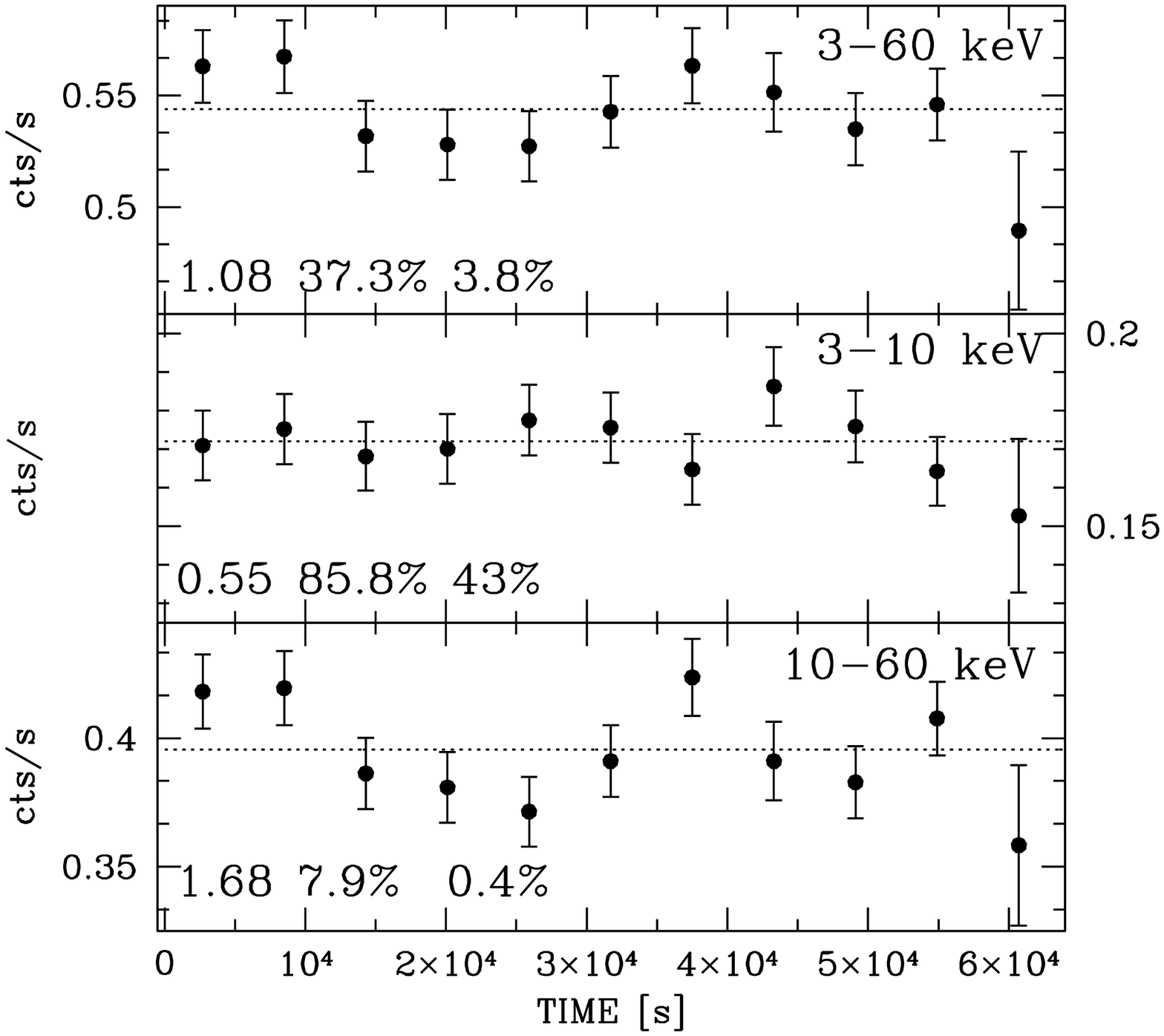}
\caption{{\it NuSTAR} light curves in bins of 5800 s ($\sim$1
  satellite orbit). The count rates are the added values detected by
  the FPMA and FPMB modules, corrected for livetime, PSF losses and
  vignetting, but not for background. The dotted lines correspond to
  the values of the count rate weighted means. The numbers in the
  lower left corner of each panel indicate the values of
  $\chi^{2}_{\nu}$ for 10 degrees of freedom, and the false positive
  $\chi^2$ and Kolmogorov Smirnov probabilities in the constant light
  curve hypothesis. From top to bottom panel: 3--60, 3--10 and
  10--60~keV count rates.}
\label{lcunu}
\end{figure}

The 3--60~keV light curve is slightly variable and shows a systematic
trend of the residuals with respect to a fit with a constant (see
Fig. \ref{lcunu}). To investigate possible spectral variations, we
analyzed the 3--10~keV and the 10--60~keV light curves separately. We
found that the 3--10~keV light curve is fully consistent with being
constant, but the 10--60~keV light curve is not, showing variability
up to $\sim$20\%. The 3--10~keV light curve is not correlated
  with the 10--60~keV at a confidence level larger than 99.99\% using
  the Spearman rank correlation coefficient. In the three energy
bands, the background level is of the order of 7\% of the total count
rate. Moreover we verified that the source and background light curves
are not correlated at a confidence level of 5\%.

\begin{figure}
\centering
\includegraphics[width=7cm]{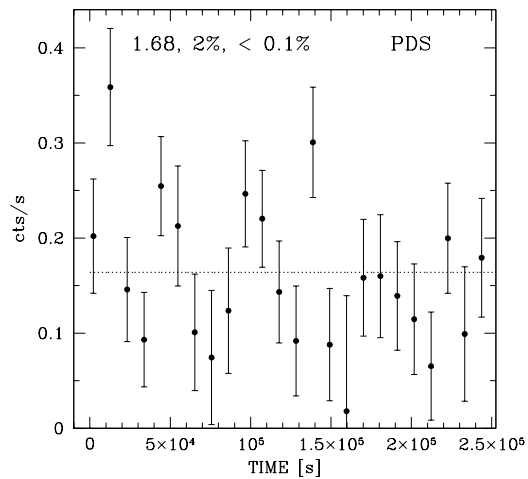}
\caption{{\it BeppoSAX} 15--200~keV PDS light curve in bins of
  10.5~ksec. The dotted line correspond to the value of the count
  rate weighted mean. The numbers in the top indicate the values of
  $\chi^{2}_{\nu}$ for 23 degrees of freedom, and the false positive
  $\chi^2$ and Kolmogorov Smirnov probabilities in the constant light
  curve hypothesis.}
\label{lcpds}
\end{figure}

To better analyze the hard X--ray variability, we analyzed {\it
  NuSTAR} light curves in three narrower energy ranges: 10--25, 25--35
and 35--60~keV. We found that the amplitude of the variations is
largest in the 25--35~keV energy band, changing by $\sim$40\% on
timescales of $\sim$15--20~ksec. We also searched for hard X--ray
variability on similarly short timescales using the archival {\it
  BeppoSAX} 15--200~keV PDS data. The PDS light curve is not constant
and the largest amplitude variability is $\sim$50\% on
timescales of $\sim$~20 ksec (see Fig. \ref{lcpds}), fully consistent
with the {\it NuSTAR} findings.

The 20--70 keV PDS mean flux, evaluated assuming a power--law model
with photon index fixed to 1.7, is a factor of $\sim$0.9 the 20--70
keV {\it NuSTAR} mean flux. This factor contains possible source
variability and also variability due to the inter--calibration
uncertainty between the two instruments.

As shown in \S 2.2, we also searched for variability in the {\it
  Chandra} data in the hardest accessible energy range (6.4--8 keV) of
the northern and the southern nuclei separately. The count rate of the
2011 ACIS--S observation of the northern nucleus is found to be
variable by a factor of $\sim 1.3$ on timescales of $\sim$~15 ksec,
while the southern nucleus is constant. In the two 2006 ACIS--S HETG
observations we did not find statistically strong evidence of
variability, neither in the southern nor in the northern nucleus. This
could be due to the poorer statistics of the zero--order ACIS--S HETG
data, in comparison to ACIS--S data (i.e. for the northern nucleus we
had 50\% fewer counts in the ACIS--S HETG).

In the 6.4--8 keV energy range, the composed light curve of the
northern and southern nuclei does not present the variability shown by
the northern nucleus alone. This is because the northern nucleus is a
factor $\sim$0.4 fainter, and the amplitude of its variations is
small. This result is consistent with the absence of variability in
the 6.4--10 keV energy range in the {\it XMM--Newton} observations.

\section{Broad--band spectral analysis}

\subsection{The model}

The extended X--ray emission at energies $<$5 keV in \textsc{NGC 6240}
is due to an absorbed multi--temperature plasma. Both the temperature
and the column density increase toward the central nuclei (Boller et
al. 2003, Netzer et al. 2005, Wang et al. 2014). The cold--temperature
plasma, which is the outermost, extends up to $\sim$1\farcm5 from the
nuclei, has sub--solar metal abundances
($\frac{Z}{Z_\odot}$$=$0.1--0.5, see Nardini et al. 2013) and
dominates at energies $\leq$1.5 keV. A medium--temperature
  plasma, which is less extended and is closer to the central
nuclei. The hot--temperature plasma, which is dominant in the
$\sim$2--5~keV energy range, indicates the presence of fast shocks
(Feruglio et al. 2013a, Wang et al. 2014, ACIS--S and HETG ACIS--S
data) and is responsible for the \ion{Fe}{XXV} line emission. In fact,
while a scattered nuclear power--law would provide an equally good
fit, the \ion{Fe}{XXV} line emission is extended and spatially
resolved by {\it Chandra} (Wang et al. 2014). Moreover a dominant
contribution to the \ion{Fe}{XXV} from photoionisation is ruled out
through the comparison of the expected X--ray line ratios with the
observed ones (Netzer et al. 2005). This is opposite to some non--LIRG
AGN (e.g. \textsc{NGC 1068}, Bauer et al. 2015 and references therein;
\textsc{Circinus}, Ar{\'e}valo et al. 2014 and references therein),
where photoionisation is dominant.

The X--ray spectrum of \textsc{NGC 6240} shows line ratios between
\ion{Mg}{XII} (1.47 keV), \ion{Si}{XIII} (1.865 keV), \ion{Si}{XIV}
(2.006 keV), \ion{S}{XV} (2.461 keV) and \ion{S}{XVI} (2.622 keV), due
to the hot thermal component, which can be hardly reproduced within
the context of equilibrium models with solar metal
abundance. Therefore, to reproduce these line ratios, the hot thermal
component was modelled by the \textsc{XSPEC}
\textsc{apec}/\textsc{mekal} models with super--solar metal abundance
($\frac{Z}{Z_\odot}$$\sim$2--2.5---Netzer et al. 2005;
$\frac{Z}{Z_\odot}$$=$10---Boller et al. 2003). A more empirical
approach was followed by Wang et al. (2014), who proposed a model with
a \textsc{mekal} component, with solar abundances, plus several
Gaussians to fit the residual lines. Alternatively, the observed
strong emission lines are expected in non--equilibrium models, in
which the ion distribution is broader than in thermal equilibrium
models with the same physical conditions; in this context, Feruglio et
al. (2013a) modelled the hot thermal component by a shock model with
almost solar abundances. Based on these arguments, we attempted to
model the hot thermal component with two different approaches: a shock
model (i.e. \textsc{XSPEC} \textsc{pshock} model, hereinafter model
A), and the usual \textsc{mekal} component (model B in the
following). As shown below, the solar chemistry can be used for model
A, whereas model B requires super--solar metal abundances.

In the energy range 5--8 keV the emission is mainly due to the two
obscured AGN in the centre of \textsc{NGC 6240} (Komossa et
al. 2003). Both nuclei are best fitted with a hard,
reflection--dominated spectrum and associated iron line complex
(i.e. Fe$K_{\alpha}$ at 6.4~keV, Fe$K_{\beta}$ at 7.06~keV rest--frame
energies and the Compton shoulder).

In the {\it NuSTAR} band, the two nuclei cannot be spatially resolved,
and their summed contribution in the hard X--ray band is modeled as
follows: a) a primary continuum, transmitted through a high column
density absorber, modeled by \textsc{plcabs} (Yaqoob 1997); b) a
reflection component, due to optically thick gas with infinite column
density illuminated by a power continuum and modeled with
\textsc{pexrav} (Magdziarz \& Zdziarski 1995); later on, we will
investigate more physically realistic reflection models. The photon
indices of the primary and the reflected components are linked to the
same value in the spectral fits. We also added three Gaussian lines to
model the Fe$K_{\alpha}$, Fe$K_{\beta}$ and Nickel K$_\alpha$ at
7.47~keV. The Fe$K_{\alpha}$ and Nickel K$_\alpha$ are significant
  at 99.99\% confidence level. The ratio between the normalization of the
Fe$K_{\beta}$ and Fe$K_{\alpha}$ is fixed to the theoretical value
(1:8.8; e.g. Palmeri et al. 2003). All the spectral components were
absorbed by the Galactic column density along the line of sight
(i.e. 4.87$\times$10$^{20}$ cm$^{-2}$; Kalberla et al. 2005) and
redshifted using $z=$0.02448 (Downes et al. 1993). We used Anders \&
Grevesse (1989) cosmic abundances. For the broad--band analysis, we
simultaneously fit the ACIS--S (0.3--8 keV), PN (2--10 keV), MOS
(MOS1$+$MOS2) (2--10 keV), FPMA (3--78.4~keV) and FPMB (3--78.4~keV)
spectra of the whole galaxy. We introduced normalization factors to
take into account the uncertain flux inter--calibration between
different instruments. We fixed the normalization factor of FPMA to
unity, leaving the other normalizations free to vary.  As discussed
above, we modeled the hot thermal components in two different
ways. Therefore we used two different models, with the following
\textsc{XSPEC} format: \\ \\ Model A:
\\ \textsc{\textsc{constant}$\times$\textsc{wabs$_{\rm
      Gal}$}$\times$(\textsc{plcabs}+\textsc{pexrav}+\textsc{zgauss}+\textsc{zgauss}+\textsc{zgauss}+
  \textsc{wabs$_{\rm cold}$}$\times$\textsc{vmekal$_{\rm cold}$} +
  \textsc{wabs$_{\rm med}$}$\times$\textsc{mekal$_{\rm med}$} +
  \textsc{wabs$_{\rm hot}$}$\times$\textsc{pshock$_{\rm hot}$})} \\

\noindent Model B:
\\
\textsc{\textsc{constant}$\times$\textsc{wabs$_{\rm Gal}$}$\times$(\textsc{plcabs}+\textsc{pexrav}+\textsc{zgauss}+\textsc{zgauss}+\textsc{zgauss}+
  \textsc{wabs$_{\rm cold}$}$\times$\textsc{vmekal$_{\rm cold}$}
  + \textsc{wabs$_{\rm med}$}$\times$\textsc{mekal$_{\rm med}$} +
  \textsc{wabs$_{\rm hot}$}$\times$\textsc{mekal$_{\rm hot}$})} \\

\noindent  where the subscripts {\rm cold}, {\rm med}, {\rm hot} indicate the cold, medium and hot--temperature components, respectively.

\subsection{Spectral analysis results}

For both models A and B, the cold thermal component requires a
sub--solar abundance ($\frac{Z}{Z_\odot}$$=$0.2--0.5) of the $\alpha$
elements (O, Ne, Mg, Si, S, Ar) and a lower sub--solar iron abundance
($\frac{Fe}{Fe_\odot}$$=$0.1--0.2) at 99.97\% confidence level
according to an F--test. This is consistent with Nardini et
al. (2013). The absorbing column density for the cold thermal
component ($N_{\rm H}=(0.12^{+0.07}_{-0.06}$)$\times$ 10$^{22}$
cm$^{-2}$, for model A) is slightly higher than, but consistent with,
the Galactic value (i.e. $N_{\rm H}=0.05$ $\times$ 10$^{22}$
cm$^{-2}$). The column densities of the gas obscuring the medium
($kT$$\sim$0.9~keV) and the hot thermal components are similar, thus
are linked together in the final fits ($N_{\rm
  H}=(0.97^{+0.18}_{-0.09}$) $\times$ 10$^{22}$ cm$^{-2}$, for model
A). This is higher than the column density of the cold thermal
component. Model B requires super--solar metal abundances for the
medium and hot--temperature components ($\frac{Z}{Z_\odot}$$\sim$2).

\begin{figure}
\begin{center}
\includegraphics[width=9.5cm]{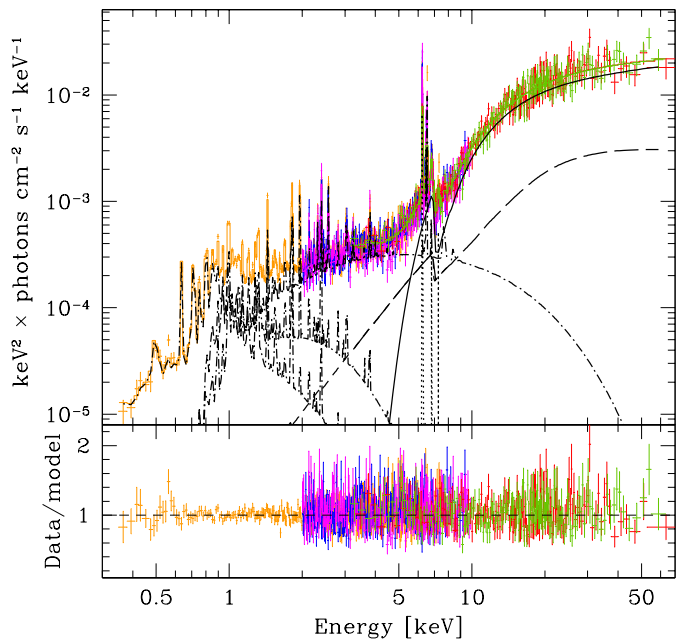}
\caption{The broad--band spectra of the whole \textsc{NGC 6240}
  galaxy, best--fit model (upper panel) and residuals (lower
  panel). The {\it NuSTAR} (green is FPMA, red is FPMB), {\it
    XMM--Newton} (blue is PN, magenta is MOS1+MOS2) and {\it Chandra}
  (orange) spectra unfolded with the instrument responses (for single
  {\it XMM--Newton} and {\it Chandra} spectra see Fig. \ref{spexmm}
  and \ref{specha}, respectively) are plotted in the upper panel. The
  total model A is superimposed. The model single continuum components
  are also shown: the primary continuum (black solid line), the
  reflection continuum (black dashed line), the Fe$K_{\alpha}$,
  Fe$K_{\beta}$ and Nickel lines (black dotted lines) and the three
  thermal components (black dashed--dotted lines). The lower panel
  shows the data/best--fit model ratios.  }
\label{speratio}
\end{center}
\end{figure}

The best--fit parameters are reported in Table~\ref{tabfit}. Both
models are fully consistent with the data ($\chi^{2}_{\nu}\sim$0.99),
nevertheless the quality of the fit for model B is slightly worse than
for model A ($\Delta \chi^2$=$+$13 for the same number of d.o.f.).
The super--solar metallicity of the hot thermal component required by
model B may indicate that the model is oversimplified. Indeed
multi--temperature plasma, non--thermal plasma and X--ray binaries
could contribute, with different percentages, to this component (see
e.g. Bauer et al. 2015; Ar{\'e}valo et al. 2014). On the other hand,
the temperature of the shock in model A
($kT$$=$$7.3^{+1.8}_{-1.4}$~keV) is slightly higher than the
temperature obtained by fitting only the {\it XMM--Newton} spectrum
($kT$$=$5.5$\pm 1.5$ keV; Boller et al. 2003, Netzer et al. 2005) or
the {\it Chandra} data ($kT$$=$6.15$\pm 0.3$ keV; Wang et al. 2014)
with similar \textsc{xspec} models, but is consistent within the
uncertainties. In fact, if the shock temperature is fixed at 6 keV the
best--fit solution is still statistically acceptable. On the basis of
the above described physical arguments, we chose model A as the
reference model. The broad--band best--fit spectra residuals and model
A are shown in Fig. \ref{speratio}.

\begin{table}
\scriptsize
\caption{Best--fitting parameters}
\begin{tabular}{p{2.2cm}p{1.3cm}p{1.3cm}p{2.2cm}}
\hline
Parameter& Model A & Model B & Units  \\ [1ex]
\hline
$N_{\rm H}$$_{\rm Gal}$$^a$&  0.0487 &    0.0487   &  10$^{22}$
cm$^{-2}$ \\ [1ex]
\hline
\end{tabular}
\begin{tabular}{l}
Primary and reflected components: \textsc{plcabs}+\textsc{pexrav}+\textsc{zgauss}+\textsc{zgauss}+\textsc{zgauss} \\ [1ex]
\end{tabular}
\begin{tabular}{p{2.2cm}p{1.3cm}p{1.3cm}p{2.2cm}}
\hline
$N_{\rm H}$$_{\rm abs}$$^b$&1.58$_{-0.09}^{+0.10}$ &         1.63 $\pm{0.10}$ &  10$^{24}$
cm$^{-2}$\\ [1ex]
$\Gamma$$^c$ &1.75$\pm{0.08}$ &    1.82 $\pm{0.07  }$ \\ [1ex]
Norm$_{\rm plcabs}$$^d$ &136$_{-36}^{+51}$ &    156$_{-41 }^{+58}$ & 10$^4$ ph keV$^{-1}$
cm$^{-2}$s$^{-1}$ \\ [1ex]
Norm$_{\rm pexrav}$$^e$ &16$\pm{6}$ &   36  $_{-6}^{+7}$ & 10$^{4}$ ph keV$^{-1}$
cm$^{-2}$s$^{-1}$\\ [1ex]
Fe$K_{\alpha}$ Energy$^f$&6.399$_{-0.007}^{+0.006}$ &  6.399 $_{-0.007 }^{+0.006}$ & keV \\ [1ex]
Fe$K_{\beta}$ Energy$^f$&7.00$\pm{0.04}$ &   7.01  $_{-0.06   }^{+0.04  }$ & keV\\ [1ex]
Nickel Energy$^f$&7.48$\pm {0.06}$ &   7.48$_{-0.06}^{+0.05}$ & keV\\ [1ex]
Fe$K_{\alpha}$ EW$^g$ &       0.33$_{-0.08}^{+0.11}$ &        0.31$_{-0.08}^{+0.10}$ & keV\\ [1ex]
Fe$K_{\beta}$ EW$^g$ &      0.04$\pm{0.01}$ &        0.04$\pm{0.01}$ & keV\\ [1ex]
Nickel  EW$^g$ &      0.09$_{-0.05}^{+0.07}$ &        0.11$_{-0.06}^{+0.08}$ & keV\\ [1ex]
\hline
\end{tabular}
\begin{tabular}{p{10cm}}
Cold--temperature component:  \textsc{wabs$_{\rm cold}$}$\times$\textsc{vmekal$_{\rm cold}$} \\ [1ex]
\end{tabular}
\begin{tabular}{p{2.2cm}p{1.3cm}p{1.3cm}p{2.2cm}}
\hline
$N_{\rm H}$$_{\rm cold}$$^h$&0.07$_{-0.06}^{+0.07}$ & 0.10  $\pm{0.05}$ & 10$^{22}$ cm$^{-2}$ \\ [1ex]
$kT$$_{\rm cold}$$^i$ & 0.59$_{-0.12}^{+0.04}$ & 0.61  $_{-0.04   }^{   +0.03  }$ & keV \\ [1ex]
ab$_O$$_{\rm cold}$$^l$ &0.35$_{-0.14}^{+0.22}$ &  0.33  $_{    -0.11   }^{    +0.18  }$ \\ [1ex]
ab$_{Fe}$$_{\rm cold}$$^l$ & 0.17$_{-0.06}^{+0.09}$ &  0.17  $_{ -0.05   }^{   +0.08  }$ \\ [1ex]
Norm$_{\rm cold}$$^m$ & 3.7$_{-1.8}^{+3.9}$ &   4.8 $_{ -2.0   }^{ +2.9  }$ & 10$^{4}$  ph keV$^{-1}$
cm$^{-2}$s$^{-1}$ \\ [1ex]
\hline
\end{tabular}
\begin{tabular}{l}
Medium--temperature component:  \textsc{wabs$_{\rm med}$}$\times$\textsc{mekal$_{\rm med}$} \\ [1ex]
\end{tabular}
\begin{tabular}{p{2.2cm}p{1.3cm}p{1.3cm}p{2.2cm}}
\hline
$N_{\rm H}$$_{\rm med}$$^h$& 0.97$_{-0.09}^{+0.18}$ &  1.2  $\pm{0.1 }$ & 10$^{22}$ cm$^{-2}$\\ [1ex]
$kT$$_{\rm med}$$^i$ &0.85$_{-0.06}^{+0.14}$ & 0.96  $\pm{   0.06  }$ & keV \\ [1ex]
ab$_{\rm med}$$^l$ &  1(fixed)  & 2.1    $_{  -0.3     }^{  +0.4    }$ \\ [1ex]  
Norm$_{\rm med}$$^m$ &6.5$_{-1.6}^{+1.4}$ &    6.1  $_{ -1.1   }^{ +1.3  }$ &10$^{4}$  ph keV$^{-1}$
cm$^{-2}$s$^{-1}$ \\ [1ex]
\hline
\end{tabular}
\begin{tabular}{lll}
Hot--temperature: & \textsc{wabs$_{\rm hot}$}$\times$\textsc{pshock$_{\rm hot}$} & \textsc{wabs$_{\rm hot}$}$\times$\textsc{mekal$_{\rm hot}$} \\ [1ex]
\end{tabular}
\begin{tabular}{p{2.2cm}p{1.3cm}p{1.3cm}p{2.2cm}}
\hline
$N_{\rm H}$$_{\rm hot}$$^h$& $=$$N_{\rm H}$$_{\rm med}$ & $=$$N_{\rm H}$$_{\rm med}$  & 10$^{22}$ cm$^{-2}$\\ [1ex]
$kT$$_{\rm hot}$$^i$ & 7.3$_{-1.4}^{+1.8}$ &   3.1    $_{  -0.3    }^{  +0.5   }$ & keV \\ [1ex]
ab$_{\rm hot}$$^l$ & $=$ab$_{\rm med}$ & $=$ab$_{\rm med}$ \\ [1ex] 
Norm$_{\rm hot}$$^m$ &7.1$\pm{0.5}$ &   6.1 $\pm{  0.8 }$ & 10$^{4}$ ph keV$^{-1}$
cm$^{-2}$s$^{-1}$\\ [1ex]
$\tau$$_u$$_{\rm hot}$$^n$ &4.4$_{-0.9}^{+1.4}$  &  & 10$^{11}$  s cm$^{-3}$\\ [1ex]
\hline
FPMB/FPMA$^o$ &1.01$\pm{0.03}$ & 1.01  $_{  -0.03    }^{  +0.04   }$ \\ [1ex]
FPMA/PN$^o$ &0.88$\pm{0.04}$ &  0.88   $\pm{  0.04   }$ \\ [1ex]
FPMA/MOS$^o$ &0.94$_{-0.04}^{+0.05}$ &  0.94   $_{ -0.04    }^{  +0.05   }$ \\ [1ex]
FPMA/ACIS--S$^o$& 1.00$_{-0.04}^{+0.05}$ & 1.00   $_{   -0.04    }^{  +0.05  }$ \\ [1ex]
$\chi^2$/d.o.f.  & 1016.2/1035   &  1028.8/1035  \\ [1ex]                               
$\chi^{2}_{\nu}$/$\chi^2$  prob. & 0.982/66$\%$  &0.994/55$\%$  \\ [1ex]    
\hline
Luminosity$^p$ \\ [1ex]
\hline
plcabs (2--10 keV) &  7.0$_{-1.9 }^{ +2.6 }$ & 7.1  $_{ -2.7}^{  +1.9 }$  & 10$^{43}$ erg s$^{-1}$\\ [1ex]
plcabs (10--40 keV) &  8.6$_{-2.3 }^{+3.3 }$ &  7.9 $_{-3.0  }^{ +2.1}$ & 10$^{43}$ erg s$^{-1}$\\ [1ex] 
plcabs (2--78 keV) &  21.8$_{-5.7}^{+8.2}$ &  20.5 $_{-5.4}^{+7.6 }$ & 10$^{43}$ erg s$^{-1}$\\ [1ex] 
pexrav (2--10 keV)&  0.06 $\pm{  0.02 }$ &  0.11 $\pm{  0.02}$ & 10$^{43}$ erg s$^{-1}$ \\ [1ex]
pexrav (10--40 keV)  & 0.64 $\pm{  0.21  }$ &1.06 $_{  -0.21  }^{+0.19}$ & 10$^{43}$ erg s$^{-1}$\\ [1ex]
$kT$$_{\rm hot}$ (2--10 keV)  &0.14 $_{-0.02  }^{  +0.01  }$ & 0.09$\pm{  0.01}$ & 10$^{43}$ erg s$^{-1}$\\ [1ex]
\hline
\end{tabular}

Best--fit values with uncertainties at the 90\% confidence level for
one parameter of interest ($\Delta$$\chi^2$=2.706) for model A and B.

$^a$ Fixed Galactic Column density $N_{\rm H}$; $^b$ column density
$N_{\rm H}$ absorbing the direct power-law; $^c$ direct power--law
photon index; $^d$ normalization at 1 keV of the direct power--law;
$^e$ normalization at 1 keV of the reflection component; $^f$
rest--frame line energy; $^g$ line equivalent width; $^h$ column
density; $^i$ plasma temperature; $^l$ metal abundances; $^m$
normalization of the thermal components; $^n$ upper limit on
ionization timescales. The subscripts {\rm cold}, {\rm med}, {\rm
    hot} in $h$, $i$, $l$, $m$ and $n$ indicate the cold, medium and
  hot--temperature components, respectively (see \S~4.1). $^o$
Normalization factors with respect to the FPMA spectrum; $^p$
intrinsic luminosities.

\label{tabfit}
\end{table}

The best--fit parameters of the absorbed power--law, describing the
nuclear X--ray continuum emission of the two nuclei, are:
$\Gamma=1.75\pm0.08$ and $N_{\rm H}= (1.58_{-0.09}^{+0.10}) \times
10^{24}$ cm$^{-2}$. The nuclear absorption--corrected 2--10 keV
luminosity is $(7.0_{-1.9 }^{+2.6 })\times 10^{43}$ erg s$^{-1}$;
the reflection component in the 10--40 keV energy range is of the
order of 20$\%$ of the observed primary emission.

We underline, as pointed out by Murphy \& Yaqoob (2009), that
combining \textsc{pexrav} and \textsc{plcabs} may produce a bias
toward solutions dominated by the direct continuum. Moreover, a
disc/slab geometry with an infinite column density for the material
responsible for the Compton--scattered continuum is assumed in the
\textsc{pexrav} model. The spectral features associated with
scattering (e.g. Fe$K_{\alpha}$ and Fe$K_{\beta}$ emission lines and
Compton shoulder) in a more realistic, finite column density medium
cannot be reproduced by infinite slab models.  The \textsc{mytorus}
model\footnote{http://www.mytorus.com/} (Murphy \& Yaqoob 2009; Yaqoob
\& Murphy 2011, and see \S~5) returns a more physically
self--consistent description of the direct continuum and the
reflection continuum and lines (Yaqoob \& Murphy 2009, 2011, Yaqoob
2012) and it should not be affected by bias toward transmitted
dominated solutions. Nevertheless, the presence in \textsc{NGC 6240}
of two nuclei, unresolved by {\it NuSTAR}, makes this source not the
best case for using models like \textsc{mytorus}. Therefore, for the
whole galaxy we used \textsc{mytorus} (described in the following
section) for a first--order comparison with the spectral results
obtained with model A.

Replacing \textsc{plcabs}$+$\textsc{pexrav} with \textsc{mytorus} in
model A, the best--fitting parameters of the absorbed power--law,
describing the nuclear X--ray continuum emission of the two nuclei,
are consistent with those obtained using model A
(i.e. $\Gamma=1.70\pm0.09$ and $N_{\rm H}=(1.39^{+0.09}_{-0.08})
\times$ 10$^{24}$ cm$^{-2}$). Moreover, the source is still not
reflection--dominated; we find that the 10--40 keV reflection
component is less than 20$\%$ of the observed direct
component. Finally, the nuclear absorption corrected 2--10~keV
luminosity is $\sim$8$\times 10^{43}$ erg s$^{-1}$, fully consistent
with model A.

\section{Resolving the two nuclei}

The \textsc{NGC 6240} hard X--ray emission is due to two AGN in the
centre of the galaxy, resolved by {\it Chandra} (Komossa et al. 2003)
but not by {\it NuSTAR} and {\it XMM--Newton}.  This is the first
attempt to disentangle the primary emission of each nucleus by
combining the spatially resolved {\it Chandra} spectra with {\it
  NuSTAR} high--energy observations.

\subsection{{\it Chandra} spectral analysis of the two individual nuclei} 

For this aim, we independently fitted the {\it Chandra} spectra of the
two nuclei (see \S~2.2) in the 0.3--8 keV energy range. 
The direct and the reflection components (continuum and iron line
emission) were modeled using the \textsc{mytorus} model (Murphy \&
Yaqoob 2009; Yaqoob \& Murphy 2011).

 Assuming that the direct continuum emission is a power--law, three
 tables are needed to generate spectral fits: \textsc{mytorusZ},
 \textsc{mytorusS}, and \textsc{mytorusL}. \textsc{mytorusZ} is a
 multiplicative table that contains the pre--calculated transmission
 factors that distort the direct continuum at all energies owing to
 photoelectric absorption and Klein--Nishina scattering (see \S~5.2 of
 the \textsc{mytorus} manual). \textsc{mytorusS} and \textsc{mytorusL}
 represent the Compton--scattered/reflected continuum toward the line
 of sight and the fluorescence emission lines typically produced in
 highly obscured AGN (i.e. Fe$K_{\alpha}$, Fe$K_{\beta}$ and Compton
 shoulder), respectively. We used the \textsc{mytorus} model in the
 original geometry (i.e. the so--called ``coupled'' mode, which
 represents a uniform torus with a half--opening angle of
 $60^{\circ}$, corresponding to a covering factor of 0.5).  The
 normalization of the line component (\textsc{mytorusL}) is linked to
 that of the scattered/reflected continuum (\textsc{mytorusS}).  The
 \textsc{mytorusL} line tables are made with a range of energy offsets
 for best--fitting the peak energies of the emission lines. We found
 that an offset of $+$30 eV is optimal for the northern nucleus. This
 energy offset is consistent with the results obtained fitting the
 spectra of the northern and the southern nucleus in the 5.5--8~keV
 energy range with a oversimplified model (i.e. empirical power--law
 continuum model plus redshifted gaussian lines). The Fe$K_{\alpha}$
 energy best--fit values are ($6.39\pm0.01$)~keV and
 ($6.42\pm0.01$)~keV, for the southern and northern nucleus,
 respectively. These values are consistent with those obtained by Wang
 et al. (2014, see their Table 2). The Fe$K_{\alpha}$ energy of the
 northern nucleus line is systematically higher than the theoretical
 value, the energy blueshift is $20_{-18}^{+14}$ eV (conservatively
 the uncertainties are at the 90\% confidence level for two parameters
 of interest, i.e. $\Delta \chi^2$=4.61). This offset could take into
 account any residual offset due to a blend of neutral and mildly
 ionized iron. Alternatively, the shift could be the Doppler effect
 due to outflowing reflecting/absorbing medium with velocity
 $-940_{-840}^{+660}$ km s$^{-1}$ (see e.g. Elvis 2000, Risaliti et
 al. 2002) or residual relative motion between the two nuclei at the
 late stage of the merging process (see e.g. Bournaud et al. 2011). In
 the latter scenario, assuming the above velocity the coalescence will
 be in ($0.7_{-0.3}^{+4.6}$) Myr, which is shorter than the timescales
 of 10-20 Myr estimated by other authors (see e.g. Scoville et
 al. 2014).

The spectra of each nucleus contain part of the X--ray emission
  of the circumnuclear plasma, because the hottest plasma probably
  originates near the nuclei. Moreover, for projection effects, the
  spectra can also be contaminated by the coldest plasma. Therefore,
  we include the multi--temperature plasma, modeled using the
  cold--temperature, the medium--temperature and the shock components
  as in model A (\S~4.2), in the spectral modelling.

The \textsc{xspec} format for the adopted model  for each nucleus is: \\ 

\textsc{zpowerlaw $\times$ \textsc{mytorusZ}($\theta_{\rm obs}$) + \textsc{mytorusS}($\theta_{\rm obs}$) +\textsc{mytorusL}($\theta_{\rm obs}$)+ \textsc{wabs$_{\rm cold}$}$\times$\textsc{vmekal$_{\rm cold}$} + \textsc{wabs$_{\rm med}$}$\times$\textsc{mekal$_{\rm med}$} + \textsc{wabs$_{\rm hot}$}$\times$\textsc{pshock$_{\rm hot}$}} 
 \\

\noindent where $\theta_{\rm obs}$ is the inclination angle between
the torus polar axis and the observer's line of sight and the photon
index of the primary continuum was fixed to the best--fit value of the
whole galaxy spectrum (see \S~4.2 and and Table~\ref{tabfit}).

We performed a fit initially fixing the parameters describing the
temperatures, absorption column densities and metal abundances of the
three thermal components, at the best--fit values obtained for the
whole galaxy spectrum (see Table~\ref{tabfit}), leaving the
normalizations free to vary. Note that the normalization of the
cold--temperature component for the northern nucleus is consistent
with zero, while for the southern nucleus it is only $\sim$4\% of the
normalization of the medium--temperature component.  We do not find a
statistically good description of the spectra
($\chi^{2}_{\nu}$$\sim$2), and some residuals are left at low
energies. The best fit for the southern nucleus
($\chi^{2}_{\nu}$/$\chi^2$ prob.$=$1.2/7$\%$) is obtained with
slightly higher values of the temperature of the medium--temperature
plasma component and the associated absorbing column density
($kT$$=(1.3\pm 0.2)$ keV, $N_{\rm H}=(1.5\pm 0.2)\times$ 10$^{22}$
cm$^{-2}$), as well as a higher value of the absorbing column density
of the shock component ($N_{\rm H}=(10\pm 3)\times$ 10$^{22}$
cm$^{-2}$). This is also the case for the northern nucleus, in which
the best--fit ($\chi^{2}_{\nu}$/$\chi^2$ prob.$=$1.3/3.4$\%$) value of
the absorbing column density of the shock component is slightly higher
($N_{\rm H}=(1.9\pm 0.3)\times$ 10$^{22}$ cm$^{-2}$) than that
obtained in \S~4.2. These results are consistent with the fact that
the extended X--ray emission of \textsc{NGC 6240}, over a large scale,
suffers from blending of thermal plasma with different temperatures,
inhomogeneous absorption and metal abundance.  Therefore, spatially
resolved analysis in small regions can give results slightly different
from the results obtained using data from larger regions (e.g. Wang et
al. 2014).

The intrinsic 2--10 keV flux due to the shock component is a factor of
$\sim$6.5 larger in the southern nucleus than in the northern,
consistent with the scenario that the shock originates from the
southern nucleus (see Feruglio et al. 2013b, Wang et al. 2014). We
also found that about 60\%, 10\% and 30\% of the total \ion{Fe}{XXV}
line flux is emitted in the southern nucleus, northern nucleus and
extended region, respectively. These percentages are consistent with
those found by Wang et al. (2014) through spectral and imaging
analysis.

\begin{figure}
\includegraphics[width=6cm,angle=-90]{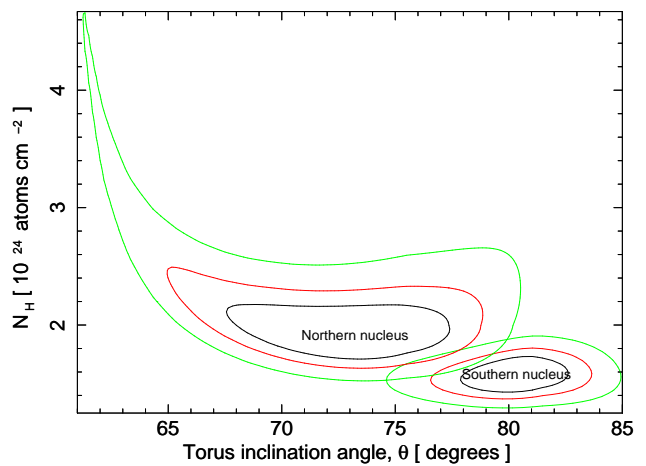}
\caption{{\it Chandra} 68\%, 90\% and 99\% confidence contours for the
  southern and northern nucleus: equatorial column density $N_{\rm H}$
  (in units of 10$^{24}$ cm$^{-2}$) versus $\theta_{\rm obs}$
  (i.e. inclination angle between the torus polar axis and the
  observer's line of sight). The confidence contours are evaluated
  leaving $N_{\rm H}$, $\theta_{\rm obs}$ and the normalizations of
  the direct continuum free to vary.}
\label{contour}
\end{figure}

The {\it Chandra} best--fit parameters of the primary continuum for
each nucleus are reported in Table~\ref{tabfit2nu}. For both nuclei,
the inclination angle of the torus and the absorbing column density
are not correlated. The inclination angle of the two tori and the
absorbing column densities are different at the $>$99\% confidence
level. The equatorial $N_{\rm H}$ is higher in the northern nucleus,
while the inclination angle is larger in the southern one (see
Fig. \ref{contour}). The actual line--of--sight column densities are:
$N_{\rm H}=(1.47^{+0.21}_{-0.17}$) $\times$10$^{24}$ cm$^{-2}$ and
$N_{\rm H}=(1.55^{+0.72}_{-0.23}$) $\times$10$^{24}$ cm$^{-2}$ for the
southern and northern nuclei, respectively. For both nuclei, this is
consistent with an optical depth to Thomson scattering of $\tau_{\rm
  T}$$\sim$1.2.\footnote{$\tau_{\rm T} =x \times \sigma_{\rm T} \times
  N_{\rm H}$, where $\sigma_{\rm T}$ is the Thomson cross section, $x$
  is the mean number of electrons per H atom, which is $\sim$1.2
  assuming cosmic abundance.} The equivalent widths of the
Fe$K_{\alpha}$ line (Table~\ref{tabfit2nu}) are fully consistent with
previous results (Wang et al. 2014).  The intrinsic primary 2--10~keV
luminosity of the southern nucleus ($5.2 \times 10^{43}$ erg s$^{-1}$)
is a factor of $\sim$2.6 larger than that of the northern.

\begin{table}
\scriptsize
\caption{{\it Chandra} best--fit parameters for the two nuclei}
\begin{center}
\begin{tabular}{lclcc}
\hline
Parameter& {\it S}$^a$ & {\it N}$^b$ & Units \\
\hline
$\Gamma$$^c$ &      1.75 & 1.75 \\ [1ex]       
$N_{\rm H}$$^d$ &       1.57$_{      -0.18}^{    +0.22}$ &       1.91$_{      -0.28}^{    +0.89}$ & 10$^{24}$ 
cm$^{-2}$ \\ [1ex] 
$A_{\rm Z}$$^e$     &    106$_{    -38}^{    +61}$ &   41$_{    -15}^{    +30}$ &  10$^4$ ph keV$^{-1}$ cm$^{-2}$s$^{-1}$\\ [1ex] 
$\Theta$$^f$ &  80 $_{-8}^{+6}$ &   73 $_{-9}^{+6}$ & deg.\\ [1ex]    
Fe$K_{\alpha}$ EW$^g$ &       0.37$_{-0.05}^{+0.24}$ &       0.58$_{-0.09}^{+0.70}$ & keV\\ [1ex]        
$\chi^2$/dof  & 135.2/ 112      & 90.75/68 \\ [1ex]
$\chi^{2}_{\nu}$/$\chi^2$  prob. &  1.2/6.7$\%$ & 1.34/3.4\% \\ [1ex]
$L_{A_{\rm Z}}$$^h$  (2--10 keV)     &   5.2 & 2.0  & 10$^{43}$ erg s$^{-1}$ \\ [1ex]
$L_{A_{\rm Z}}$$^i$  (10--40 keV)    &   7.1 &  2.7 & 10$^{43}$ erg s$^{-1}$ \\ [1ex]
$L_{A_{\rm Z}}$$^i$  (2--78 keV)    &   15.7 & 6.0  & 10$^{43}$ erg s$^{-1}$ \\ [1ex]
$L_{A_{\rm S}}$$^i$  (10--40 keV)    &   1.2 &  0.5 & 10$^{43}$ erg s$^{-1}$\\ [1ex]
\hline
\end{tabular}
\end{center}

{\it Chandra} best--fit values with uncertainties at the 90\%
confidence level for one parameter of interest ($\Delta
\chi^2$=2.706). 

$^a$ Best--fit parameters for the southern nucleus; $^b$ best--fit
parameters for the northern nucleus; $^c$ direct power--law photon
index fixed to the best--fit value of the total spectrum (see \S~4.2);
$^d$ equatorial column density $N_{\rm H}$; $^e$ normalization of the
primary continuum referred to 1 keV; $^f$ inclination angle between
the torus polar axis and the observer's line of sight; $^g$ line
equivalent width; $^h$ intrinsic primary 2--10~keV luminosities; $^i$
intrinsic 10--40~keV and 2--78~keV luminosities evaluated extrapolating the {\it
  Chandra} best--fit model using the {\it NuSTAR} response efficiency:
$L_{A_{\rm Z}}$ for the primary continuum, $L_{A_{\rm S}}$ for the
reflected/scattered component.

\label{tabfit2nu}
\end{table}

\subsection{{\it Chandra} model of the two nuclei versus the {\it NuSTAR} spectrum} 

\begin{figure}
\begin{center}
\includegraphics[width=8cm]{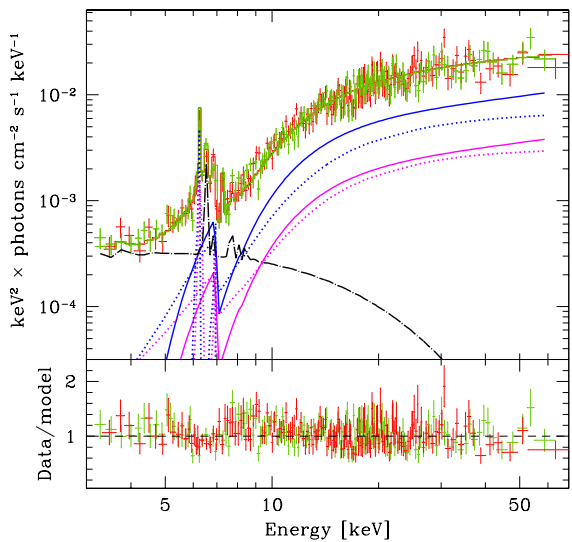}
\caption{ The {\it NuSTAR} spectra unfolded with the instrument
  responses (green is FPMA, red is FPMB) superimposed with the
  best--fit thermal component of the whole galaxy (black long
  dash--dotted line, medium--temperature and shock components in model
  A: \textsc{wabs$_{\rm med}$}$\times$\textsc{mekal$_{\rm med}$} +
  \textsc{wabs$_{\rm hot}$}$\times$\textsc{pshock$_{\rm hot}$}, see
  \S~4.2) and the {\it Chandra} best--fit \textsc{mytorus} model for
  the two nuclei (see \S~5). The blue lines mark the primary continuum
  (blue solid line), and the reflection continuum and lines (dotted
  line) of the southern nucleus; the magenta lines are the primary
  continuum (magenta solid line), the reflection continuum and lines
  (magenta dotted line) of the northern nucleus. In the lower panel
  the {\it NuSTAR} data/{\it Chandra} best--fit model ratios are
  shown.  }
\label{speratio2nuc}
\end{center}
\end{figure}

The sum of the best--fit models of the {\it Chandra} spectra of the
two nuclei described above is extrapolated to high energies and
over--plotted on the {\it NuSTAR} data of the whole galaxy in
Fig. \ref{speratio2nuc}. A statistically acceptable description of
the {\it NuSTAR} spectrum is obtained without adjusting the fit:
$\chi^{2}_{\nu}$/$\chi^2$ prob.$=$1.16/2.7$\%$ for the 3--70 keV
energy range, and 1.09/15$\%$ ignoring the 7.5--9 keV energy range,
where the discrepancies are the largest as judged by the residuals in
the lower panel of Fig. \ref{speratio2nuc}.

Leaving the normalization of the direct continuum and the equatorial
$N_{\rm H}$ for the southern or the northern nucleus free to vary, the
goodness of the fit improves at 99.8\% confidence level. While the
best--fit values of the normalization of the direct continuum remain
consistent with those previously measured, the best--fit values of the
column density are systematically lower ($N_{\rm
  H}$$=$$(1.38\pm{0.06}) \times 10^{24}$~cm$^{-2}$ and $N_{\rm
  H}$$=$$(1.33_{-0.15}^{+0.16}) \times 10^{24}$~cm$^{-2}$ for the
southern and northern nucleus, respectively). This indicates possible
column density variations $\leq 2 \times 10^{23}$~cm$^{-2}$ on timescales
of$\sim$3~years. Fits were also performed leaving the spectral slope
free to vary. While for the southern nucleus the best--fit photon
index is relatively well constrained ($\Gamma=1.70\pm0.08$), only a
lower limit could be obtained for the northern one ($\Gamma >
1.67$). In any case, these slopes are fully consistent with that
obtained fitting the total spectrum ($\Gamma=1.75$).

\section{Discussion}

The two nuclei in \textsc{NGC 6240} are separated by $\sim1\farcs5$
($\sim$0.7 kpc, Fried \& Schultz 1983) and are therefore not spatially
resolved by {\it NuSTAR} or {\it XMM--Newton}. As a first step, we
performed joint {\it NuSTAR}$+${\it XMM--Newton}$+${\it Chandra}
broad--band spectral analysis (0.3--70 keV) of the data of the whole
\textsc{NGC 6240} galaxy (i.e. combination of the two nuclei, extended
emission and serendipitous sources), to evaluate the mean properties
of the whole system and for comparison with previous high energy
observations.

The broad--band 0.3--70 keV spectrum is characterized by three main
spectral components (model A in \S 2.1 and Table~\ref{tabfit}). The
first is the primary continuum emission piercing through a
Compton--thick obscuring medium which dominates the spectrum at
energies $>$~10~keV, where most of the X-ray luminosity is
observed. The second component is the reflected/scattered emission
from cold gas which also generates strong Fe$K_{\alpha}$ and
Fe$K_{\beta}$ emission lines at 6--7 keV.  The third is a
multi--temperature thermal plasma due to the nuclear starburst and
extended super--wind.  Both the transmitted and the reflected
components are required by the best fit at a high level of
significance ($> 5\sigma$) according to an F--test.  The intensity of
the reflection component in the 10--40 keV energy range is of the
order of, or slightly lower, than $\sim$20$\%$ of the observed primary
emission. Therefore, most of the high energy flux is transmitted,
rather than Compton--scattered, and the whole system is not
reflection--dominated. This is consistent with the relatively low
value of the Fe$K_{\alpha}$ line equivalent width ($\sim$0.3~keV; see
also Shu et al. 2011, Brightman \& Nandra 2011) and the observed
variability at energies $>$10 keV.

The nuclear emission is characterized by a primary power law continuum
with $\Gamma=1.75\pm 0.08$ and $N_{\rm H}=
(1.58_{-0.09}^{+0.10})\times$10$^{24}$ cm$^{-2}$, and intrinsic
  luminosity of ($7.0_{-1.9 }^{+2.6 })\times$10$^{43}$ erg s$^{-1}$
  and of ($21.8_{-5.7 }^{+8.2})\times$10$^{43}$ erg s$^{-1}$ in the
  2--10 keV and 3--78 keV energy ranges, respectively. Our findings
significantly improve upon previous {\it BeppoSAX} (Vignati et
al. 1999) and {\it RXTE} (Ikebe et al 2000) results, both for
disentangling the primary and reflection components, and for better
constraining the spectral parameters.

The {\it BeppoSAX} best--fit absorbing column density ( $N_{\rm H}=
(2.18_{-0.27}^{+0.30})\times 10^{24}$ cm$^{-2}$) is higher than the
{\it NuSTAR} best fit value at $> 3 \sigma$. To better investigate
this discrepancy we re--analyzed the {\it BeppoSAX} data using our
best--fit model A (see \S 2.1 and Table~\ref{tabfit}) and leaving only
the column density obscuring the primary continuum free to vary. We
introduced a normalization factor between our best--fit model A and
the {\it BeppoSAX} data, to take into account the flux
inter--calibration and the contribution from the serendipitous sources
in the MECS extraction region (radius 4$\amin$, see \S~2.4). The
best--fit cross--normalization factor is 1.09$\pm0.05$.  We also added
a normalization factor between MECS and PDS data following the ABC
{\it BeppoSAX} data analysis guide.\footnote{http://heasarc.nasa.gov/docs/sax/abc/saxabc/saxabc.html} The
best--fit ($\chi^{2}_{\nu}$/$\chi^2$ prob.$=$0.84/79$\%$) value of the
absorbing column density is $N_{\rm H}=(1.82_{-0.10}^{+0.12})\times
10^{24}$ cm$^{-2}$, which is lower, although consistent within the
errors, than the Vignati et al. (1999) best fit value. The {\it
  BeppoSAX} best--fit $N_{\rm H}$ remains systematically higher than
that of {\it NuSTAR} by about $\sim$(1--2)$\times 10^{23}$~cm$^{-2}$,
and the bolometric luminosity is $\sim 9\%$ larger than that measured
from {\it NuSTAR}.  This luminosity difference includes the
inter--calibration uncertainty between the {\it NuSTAR} and the {\it
  BeppoSAX} data, and is consistent with little intrinsic variability
in \textsc{NGC 6240}. This would also suggest that the {\it BeppoSAX}
data are not contaminated by bright serendipitous sources.

\subsection{Extended emission}

The {\it NuSTAR} data have made possible a precise measurement of the
nuclear continuum of \textsc{NGC 6240} over a broad energy range and
in turn allowed better constraints on the softer components associated
with the host galaxy.  More specifically, the medium--temperature
plasma (\textsc{mekal$_{\rm med}$}, $kT$$_{\rm med}$ in
Table~\ref{tabfit}) has an observed luminosity of $2\times 10^{41}$
erg s$^{-1}$ and $\sim 0.9 \times 10^{41}$ erg s$^{-1}$ in the 0.5--2
and 2--10~keV energy ranges, respectively. Star--forming galaxies are
luminous sources of X--ray emission, which originates from X--ray
binaries, young supernovae remnants, hot (0.2--1 keV) interstellar gas
associated to star-forming regions and O stars (Fabbiano 1989, 2006
for reviews). Their X--ray luminosity correlates with the SFR (see,
e.g. Ranalli et al. 2003; Lehmer et al. 2010 and reference therein),
thus using the \textsc{NGC 6240} SFR (61 $M_\odot$ yr$^{-1}$ Yun \&
Carilli 2002) and the relations by Ranalli et al. (2003) and Lehmer et
al. (2010), we estimate a 0.5--2~keV luminosity of $2.7\times 10^{41}$
erg s$^{-1}$ and a 2--10 keV luminosity of $0.7_{-0.4}^{+0.9} \times
10^{41}$ erg s$^{-1}$. Despite the uncertainties that affect these
relations, these results suggest that the medium--temperature
component can be fully due to star formation.

The hot temperature plasma ($kT$$_{\rm hot}$ in Table~\ref{tabfit}) is
fitted with both a non--equilibrium model (model A in
Table~\ref{tabfit}) and an equilibrium model (model B in
Table~\ref{tabfit}). Using model A, the ionization timescale parameter
is $(4.4_{-0.9}^{+1.4})\times 10^{11}$ s cm$^{-3}$. If the plasma were
close to being in collisional ionization equilibrium then we would
expect a value $\ge 10^{12}$ s cm$^{-3}$ (Masai 1994), suggesting that
the plasma is in a marginally non--equilibrium state. The
non--equilibrium model is statistically preferred over the equilibrium
one ($\Delta \chi^2$=$+$13 for the same number of d.o.f). Furthermore,
it does not require super--solar metal abundances, required for model
B ($\frac{Z}{Z_\odot}$$\sim$2). The non--equilibrium model strongly
favours the presence of a nuclear shock (Feruglio et al. 2013a, Wang
et al. 2014).

For the shock component, we evaluated a best--fit value of the column
density value of $N_{\rm H}$$\simeq 10^{23}$ cm$^{-2}$ and an
intrinsic 0.5--8~keV luminosity of $\sim 3\times 10^{42}$ erg
s$^{-1}$. This value is about a factor of 0.5 smaller than the
luminosity obtained from the {\it Chandra} spectrum only, but, as
noticed by Wang et al (2014), their absorption--corrected luminosity
could be overestimated by a factor of two, given their unlike large
column density of the gas obscuring the hot plasma (i.e. $N_{\rm H}=
(5.5\pm1.7)\times 10^{23}$ cm$^{-2}$).

\subsection{AGN contribution}

The total AGN bolometric luminosity, calculated assuming the
bolometric corrections for type 2 AGN of Lusso et al. (2011) ($\sim$15
for the observed X--ray luminosity, with an uncertainty of a factor of
$\sim$3) and the absorption--corrected nuclear 2-10 keV X--ray
luminosity, is $(1.1_{-0.3}^{+0.5})\times$10$^{45}$ erg s$^{-1}$. This
value is a factor of $\sim$2 lower than that evaluated from the
spectral energy distribution (Lira et al. 2002, $L_{bol}
\sim2\times$10$^{45}$ erg s$^{-1}$, assuming the cosmology adopted in
this paper). We note that Lira et al. (2002), extrapolating the
best--fit of Vignati et al. (1999), assumed an AGN unabsorbed
2--10~keV continuum flux of $1\times$10$^{-10}$ erg s$^{-1}$
cm$^{-2}$, which is $\sim$ twice ours; this explains the discrepancy
in the bolometric luminosity. We also note that a lower value of the
total AGN bolometric luminosity reduces the discrepancy between the
luminosity of the 140--$K$ component and the absorbed AGN continuum
found by the spectral energy distribution (Lira et
al. 2002). Therefore, \textsc{NGC 6240} is one of the most luminous
among the local bona fide Compton--Thick AGN, similar to \textsc{Mrk
  34} (Gandhi et al. 2014).

Assuming that the 8--1000 $\mu$m luminosity of~$3.4\times10^{45}$ erg
s$^{-1}$ (Wright et al. 1984; Sanders et al. 2003) is a good proxy for
the total luminosity, the two AGN contribute $\sim (32_{-9}^{+15})$\%
of the total energy output. This is consistent with the values
obtained using multiple methods such as SED$+$Infrared spectroscopy
fitting, PAH features, high/low excitation MIR lines, MIR dust
continuum (e.g., 20--24\%--Armus et al. 2006; 25--50\%--Lutz et
al. 2003; up to 60\%--Egami et al. 2006; 25.8\%--Veilleux et al. 2009;
45--60\%--Mori et al. 2014). The AGN fraction is relatively high for
the observed infrared luminosity but is not
implausible. Alonso--Herrero et al. (2012) showed that in local LIRGs
the AGN bolometric contribution to the IR luminosity of the system is
$>25\%$ in about 8\% of the sample. Moreover, in the local Universe
the AGN contribution increases with the IR luminosity of the system
(Nardini et al. 2010; Imanishi et al. 2010a,b; Alonso Herrero et
al. 2012), and \textsc{NGC 6240} is a high--luminosity LIRG.

\subsection{The accretion rate of the two nuclei}

The spatially unresolved {\it NuSTAR} spectrum is consistent with the
sum of the best--fit models obtained from the spectral analysis of the
{\it Chandra} data of the two nuclei. This suggests that the emission
at energies $>$10~keV of \textsc{NGC 6240} is due only to the two
nuclei, and is not strongly variable on long timescales
($\sim$3~years). Moreover, there are possible column density
variations of $\leq 2 \times 10^{23}$~cm$^{-2}$. The two nuclei are
both highly absorbed with $\tau_{\rm T}\sim$1.2 (i.e. $N_{\rm H}\sim
1.5\times$ 10$^{24}$ cm$^{-2}$) and at energies $>$10~keV they are
dominated by the primary continuum emission. The actual
line--of--sight column density is a few 10$^{22}$ cm$^{-2}$ larger in
the northern nucleus; this would be consistent with the results from
the 3--5~$\mu$m spectroscopy (Risaliti et al. 2006). The similarity of
the column densities of the Compton--thick material obscuring the two
nuclei could suggest that there is a common obscurer along the
line--of--sight, possibly originating in gas and dust thrown up by the
ongoing galaxy merger. The current data do not allow us to determine
where this material lies. The calorimeter onboard the {\it Astro-H}
mission (Takahashi et al. 2014) will allow to measure the profile of
the Fe$K_{\alpha}$ with a resolution of the order of
$\approx$\,350\,km\,s$^{-1}$ and thus it will be possible to
discriminate whether it is originated in a compact circumnuclear torus
or in larger scale material in the host galaxy (Gandhi et al. 2015).

The intrinsic primary 2--10~keV luminosities of the southern ($5.2
\times 10^{43}$ erg s$^{-1}$) and northern nucleus ($2 \times 10^{43}$
erg s$^{-1}$) are fairly consistent, within a factor of $\sim$2, with
those expected from the MIR--X--ray luminosity correlation (Gandhi et
al. 2009), using the nuclear 12~$\mu$m fluxes (Asmus et al. 2014).

The bolometric luminosities are 8$\times$10$^{44}$ erg s$^{-1}$ and
2.6$\times$10$^{44}$ erg s$^{-1}$ for the southern and northern
nuclei, respectively, using bolometric corrections for type 2 AGN from
Lusso et al. (2011). The southern nucleus is brighter than the
northern as expected according to the 3--5~$\mu$m luminosity (see e.g.
Risaliti et al. 2006, Mori et al. 2014). The bolometric luminosity of
the southern nucleus is lower, but consistent, with the value from the
spectral energy distribution in the NIR/MIR bands
($L_{bol}=(1.4\pm{0.6})\times$10$^{45}$ erg s$^{-1}$, assuming the
cosmology adopted in this paper). The slight, not significant,
discrepancy could be due to a residual starburst contribution in the
MIR flux.

The inferred accretion rate, $\lambda_{\rm Edd} = L_{\rm bol}/L_{\rm
  Edd}$ is $\sim$0.005$^{+0.003}_{-0.002}$ for the southern nucleus,
assuming a black hole mass of (0.84--2.2)$\times 10^{9} M_{\odot}$
(Medling et al. 2011), and $\lambda_{\rm
  Edd}$$\sim$0.014$^{+0.006}_{-0.003}$ for the northern nucleus
assuming a black hole mass of $(1.4\pm0.4)\times 10^{8} M_{\odot}$
(Engel et al. 2010). The northern nucleus shows a higher $\lambda_{\rm
  Edd}$, but the uncertainties on its black hole mass are also a
factor of a few. On the contrary, the black hole mass of the southern
nucleus is more precise, being determined by high resolution stellar
kinematics. The estimated accretion rates are on the low tail of the
distribution of Eddington ratios in nearby Seyfert 2 galaxies
(e.g. Vasudevan et al. 2010). In this respect, it is similar to the
obscured AGN selected in the GOODS survey (Simmons et al. 2011,
2012). The low value of the accretion rate and the high black hole
mass for both nuclei suggests that they could have already assembled
most of their mass through accretion processes and are likely to
become inactive. The low level of accretion in a luminous, highly
obscured early merger system is somewhat unexpected. The relatively
low accretion rate may be associated with the strong molecular outflow
(Feruglio et al. 2013b) which has depleted the nuclear regions of
gas. We also note that in spite of the large uncertainties in the
black hole mass and Eddington ratio, the photon index of the primary
continuum is consistent with the low accretion rates of the two nuclei
according to the relation found for a few AGN type 1 samples (Shemmer
et al. 2008, Risaliti et al. 2009, Brightman et al. 2013).

\subsection{Hard X--ray variability}

A detailed analysis of the light curves in several independent energy
bands clearly reveals hard X--ray variability on timescales of
$\sim$15--20~ksec by up to $\sim$20\%. The variability is peaked at
$\sim$30 keV ($\sim$40\%). These results are consistent with the
variability detected earlier in the {\it Swift} BAT data on
substantially longer timescales (i.e. months/years). In fact, Soldi et
al. (2013) found maximum (54$\pm18$\%) and minimum variability
(4$\pm3$\%) in the 14--24 and 35--100 keV energy bands, respectively.
Moreover, by re--analyzing the {\it BeppoSAX} PDS data
($\sim$15--200~keV), we found variability up to $\sim$50\% on
timescales fully consistent with those of the {\it NuSTAR} hard X--ray
light curve. The PDS has a field of view (FOV) of 1.3$^\circ$ and
therefore the data could suffer from contamination. Nevertheless, for
\textsc{NGC 6240}, no known bright sources are present in the PDS FOV,
and the probability of a serendipitous source in the PDS FOV with a
flux equal or larger than that of \textsc{NGC 6240} is $\leq 10^{-5}$
(see Vignati et al. 1999).

Assuming that the {\it NuSTAR}, {\it BeppoSAX} PDS and {\it Swift} BAT
hard X--ray variability have the same origin, the observed variability
can be explained by the two following hypotheses:

a) the variability is due to flux variations of the transmitted
component of the southern nucleus up to $\sim$20--30\% for
{\it NuSTAR}, and by $\sim$50-70\% for the {\it BeppoSAX} PDS and {\it
  Swift} BAT observed variability. Indeed, the lack of variability of
the Fe$K_{\alpha}$ line suggests that the reflection component is
constant. Moreover the spectral analysis suggests that the nuclei are
transmission dominated at energies $>$10~keV.

b) On the contrary, if the variability is due to the northern nucleus,
the amplitude of the variations should be up to $\sim$50--60\% for
{\it NuSTAR}, and up to of $\sim$2.6 for the {\it BeppoSAX} PDS and
{\it Swift} BAT observed variability. The {\it NuSTAR} variability
would be consistent with the tentative $\sim$30\% variability in the
6.4--8 keV {\it Chandra} light curve of the northern
nucleus. Nevertheless the {\it NuSTAR}, {\it BeppoSAX} PDS and {\it
  Swift} BAT hard X--ray variability would imply that the northern
nucleus should be as bright, or brighter than the southern nucleus in
some periods. This is difficult to reconcile with the evidence that
over the last $\sim$15 years, all data suggest the southern nucleus
has been brighter than the northern as indicated by {\it Chandra} data
(Komossa et al. 2003; Shu et al. 2011; Wang et al. 2014) and
3--5~$\mu$m luminosity (see e.g. Mori et al. 2014, Risaliti et
al. 2006). Therefore the most plausible hypothesis is that the hard
X--ray variability is mainly due to variations of the transmitted
component of the southern nucleus.

The 6.4--8~keV tentative variability of the northern nucleus suggests
changes of the direct continuum or/and of the column density; we are
not able to disentangle them with the current data. In any case, the
different 6.4-8 keV variability could be an indication that the matter
reflecting/Compton--scattering the nuclear emission is distributed
differently in the nuclei, as the results obtained in the present
analysis seem to suggest (see Fig.\ref{contour}).

On timescales of the order of years, the mean hard X--ray flux state
is approximately constant, suggesting that the accretion rates of the
two nuclei do not experience strong variability on year--long
timescales.

The variability observed in \textsc{NGC 6240} is similar to that
observed in \textsc{NGC 4945}. Indeed also in \textsc{NGC 4945} the
hard X--ray variability is ascribed to changes of the primary
continuum, confirmed through a detailed count rate resolved spectral
analysis (Puccetti et al. 2014). In \textsc{NGC 4945}, the fast and
large hard X--ray variability, the constancy of the Fe$K_{\alpha}$
line and the huge Compton depth along the line of sight ($\tau_{\rm
  T}\sim 2.9$) suggest a low global covering factor ($\sim$0.15) for
the circumnuclear gas. In comparison with \textsc{NGC 4945}, in
\textsc{NGC 6240} the Fe$K_{\alpha}$ line is constant, but the hard
X--ray variability is smaller and slower, likely due to the higher
black hole masses and lower accretion rates, and the Compton depth
along the line of sight is a factor $\sim$2.5 lower. These findings
indicate that the global covering factor for \textsc{NGC 6240} should
be larger than in \textsc{NGC 4945}.

\section{Conclusion}

We present {\it NuSTAR} observations of \textsc{NGC 6240}, one of the
nearest luminous infrared galaxies in a relatively early merger state,
with two distinct nuclei separated by a distance of
$\sim 1\farcs5$. The main results of this work can be summarized as
follows:

\begin{itemize}

\item We clearly detect at a high significance ($>5 \sigma$) both
  the transmitted primary absorbed continuum and the cold reflection
  continuum. The primary continuum is obscured by a $N_{\rm H}$ of
  $\sim1.6 \times 10^{24}$ cm$^{-2}$, and dominates at energies
  $>$10~keV.

\item The total AGN bolometric luminosity is
  ($1.1_{-0.3}^{+0.5})\times$10$^{45}$ erg s$^{-1}$, which is
  $(32_{-9}^{+15})$\% of the infrared luminosity. This indicates that
  the galaxy emission is not dominated by the two AGN, but likely by a
  starburst and related processes.

\item The spatially unresolved {\it NuSTAR} hard X--ray spectrum of
  \textsc{NGC 6240} is consistent with the sum of the {\it Chandra}
  best--fit models of the two nuclei extrapolated to the {\it NuSTAR}
  hard X--ray energies. This suggests that the hard emission of
  \textsc{NGC 6240} is only due to the two nuclei, and that it has not
  varied by much on a decade--long timescale. Both nuclei are highly
  absorbed with $\tau_{\rm T}\sim$1.2 (i.e. $N_{\rm H}\sim 1.5\times$
  10$^{24}$ cm$^{-2}$) as expected in the early stage of a major
  merger, and have bolometric luminosities $\sim 8 \times 10^{44}$ erg
  s$^{-1}$ (southern nucleus) and $\sim$2.6$\times 10^{44}$ erg
  s$^{-1}$ (northern nucleus). The two nuclei show low accretion rate
  ($\lambda_{\rm Edd} = L_{\rm bol}/L_{\rm Edd}$ is
  $\sim$0.005$^{+0.003}_{-0.002}$ for the southern nucleus, and
  $\lambda_{\rm Edd}$$\sim$0.014$^{+0.006}_{-0.003}$ for the northern
  nucleus).

\item We found variability in the {\it NuSTAR} light curve at energies
  $>$10~keV of $\sim$20\% on timescales of 15--20~ksec. We
  found similar variability in the {\it BeppoSAX} PDS light
  curve. Various arguments lead us to conclude that the hard X--ray
  variability is due to variations of the primary continuum of the
  southern nucleus.

\item Finally, by comparing {\it NuSTAR} and {\it BeppoSAX}
  and {\it Chandra} archival data, we found that the two nuclei
  remain consistently Compton--thick. Although we find evidence of
  variability of the material along the line of sight with column
  densities $N_{\rm H}$$\leq$2$\times$10$^{23}$ on timescales of
  $\sim$3--15~years.

\end{itemize}

Variability below 10~keV is not observed in Compton--thick AGN due to
absorption and reflection. Above 10~keV only a few sources have been
sampled with sufficient statistics, and {\it NuSTAR} is finding
variability to be more common (\textsc{NGC 4945}, Puccetti et
al. 2014; \textsc{NGC 1068}, Marinucci et al. submitted; \textsc{NGC
  6240}, this work). Therefore to investigate variability in this type
of source, good sampling of the 30--50 keV energy range is
mandatory. Further {\it NuSTAR} monitoring of \textsc{NGC 6240} would
allow a detailed time/count rate resolved spectral analysis and help
put constraints on the geometry/location of the reflecting/absorbing
medium.

\begin{acknowledgements}

This work was supported under NASA Contract NNG08FD60C, and made use
of data from the {\it NuSTAR} mission, a project led by the California
Institute of Technology, managed by the Jet Propulsion Laboratory, and
funded by the National Aeronautics and Space Administration. We thank
the {\it NuSTAR} Operations, Software and Calibration teams for
support with the execution and analysis of these observations. This
research has made use of the {\it NuSTAR} Data Analysis Software
(NuSTARDAS) jointly developed by the ASI Science Data Center (ASDC,
Italy) and the California Institute of Technology (USA).  AC, AM, FF
and LZ acknowledge support from the ASI/INAF grant I/037/12/0 --
011/13. WNB acknowledges support from Caltech NuSTAR sub-contract
44A-1092750. FEB and CR acknowledge support from CONICYT--Chile grants
Basal--CATA PFB--06/2007 and FONDECYT 1141218. FEB, CR and PA
acknowledge support from "EMBIGGEN" Anillo ACT1101. FEB acknowledges
support from the Ministry of Economy, Development, and Tourism's
Millennium Science Initiative through grant IC120009, awarded to The
Millennium Institute of Astrophysics, MAS. MB acknowledges support
from NASA Headquarters under the NASA Earth and Space Science
Fellowship Program, grant NNX14AQ07H.

\end{acknowledgements}


\end{document}